\documentclass[a4paper,11pt]{article}
\usepackage{jheppub}

\usepackage[utf8]{inputenc}
\usepackage{amsmath,amssymb}
\usepackage{amsfonts}
\usepackage{bm,bbm}
\usepackage{graphicx}
\usepackage{slashed}
\usepackage{booktabs} 
\usepackage{float}
\usepackage[dvipsnames]{xcolor}
\usepackage[most]{tcolorbox}
\usepackage{colortbl}
\usepackage[normalem]{ulem}
\usepackage{soul}
\usepackage{siunitx} 
\usepackage{hyperref}
\hypersetup{colorlinks,citecolor= blue,linkcolor= blue, urlcolor=blue}
\usepackage{ulem}
\usepackage{array}
\usepackage{verbatim}
\usepackage{epsfig}
\usepackage{multirow}
\usepackage{bbold}
\usepackage{here}
\usepackage{orcidlink}
\usepackage[version=4]{mhchem}
\usepackage[capitalise]{cleveref}
\usepackage[toc,page]{appendix}

\allowdisplaybreaks[4]
\def\lsim{\mathrel{\raise.3ex\hbox{$<$\kern-.75em\lower1ex\hbox{$\sim$}}}}
\def\gsim{\mathrel{\raise.3ex\hbox{$>$\kern-.75em\lower1ex\hbox{$\sim$}}}}

\newcommand{\Tr}{\mbox{Tr}}

\newcommand{\hc}{{\tt h.c.} }
\def\l{\left}
\def\r{\right}

\def\nn{\nonumber}
\def\l{\left}
\def\r{\right}

\title{\boldmath Effective field theory for type II seesaw model
--symmetric phase v.s. broken phase--}

\author[a,b]{Yi Liao\,\orcidlink{0000-0002-1009-5483},}
\emailAdd{liaoy@m.scnu.edu.cn}
\affiliation[a]{State Key Laboratory of Nuclear Physics and
Technology, Institute of Quantum Matter, South China Normal
University, Guangzhou 510006, China}
\affiliation[b]{Guangdong Basic Research Center of Excellence for
Structure and Fundamental Interactions of Matter, Guangdong
Provincial Key Laboratory of Nuclear Science, Guangzhou
510006, China}
\author[a,b]{Xiao-Dong Ma\,\orcidlink{0000-0001-7207-7793},}
\emailAdd{maxid@scnu.edu.cn}
\author[a,b]{and Yoshiki Uchida\,\orcidlink{0000-0003-4540-7595}}
\emailAdd{uchida.yoshiki@m.scnu.edu.cn}

\abstract{
Effective field theory is an effective approach to parameterizing effects of high energy scale physics in low energy measurements. 
The two popular frameworks for physics beyond the standard model are the so-called standard model effective field theory (SMEFT) and the Higgs effective field theory (HEFT). 
While the description by the SMEFT deteriorates when the scale of new physics is not so high or it participates in spontaneous electroweak symmetry breaking, the HEFT makes use of nonlinear realization of spontaneously broken symmetry in which there are practically no restrictions on the Higgs field as a singlet. 
In this work we present another framework, called broken phase effective field theory (bEFT), in which we deal directly with mass eigenstate fields after spontaneous symmetry breaking without employing nonlinear realization. 
We take the type-II seesaw model as an example to demonstrate our approach. By matching the model to both the bEFT and the SMEFT at tree level, we compare the results for two processes, the Higgs pair production via vector boson fusion which appears as a subprocess at the LHC and the Higgs-strahlung process at a future electron-positron collider. 
We find that the bEFT reproduces the type-II seesaw model more accurately than the SMEFT in the regions where the bare mass of the Higgs triplet becomes close to the electroweak scale. 
Therefore, the bEFT serves as a useful framework that can compensate for the shortcomings in both the SMEFT and the HEFT when dealing with UV models that involve Higgs mixing or new particles with a mass close to the electroweak scale.
}

\makeatletter
\gdef\@fpheader{}
\makeatother
\begin{document} 

\maketitle
\setcounter{page}{2}

\section{Introduction
\label{sec:intro}}

With the discovery of the Higgs boson at the Large Hadron Collider (LHC) in 2012~\cite{ATLAS:2012yve,CMS:2012qbp}, the standard model (SM) was finally established. 
However, there are some problems that the SM cannot address, such as the origin of neutrino mass. 
In order to solve this problem, several types of new physics (NP) models have been proposed, such as the three conventional seesaw models \cite{Minkowski:1977sc, Yanagida:1979as, Gell-Mann:1979vob, Glashow:1979nm, Mohapatra:1979ia, Shrock:1980ct, Schechter:1980gr, Konetschny:1977bn, Magg:1980ut, Cheng:1980qt, Mohapatra:1980yp, Lazarides:1980nt, Foot:1988aq} and radiative models aiming at a common origin of neutrino mass and dark matter~\cite{Tao:1996vb,Ma:2006km}; for a review see, e.g., Ref.~\cite{Cai:2017jrq}. 
Usually the new particles are much heavier than the electroweak scale, securing an effective field theory (EFT) approach to their effects in low energy physical processes.

There are two representative EFTs to parameterize the deviations caused by heavy NP from the SM predictions. 
One is the standard model effective field theory (SMEFT)~\cite{Weinberg:1979sa,Buchmuller:1985jz,Leung:1984ni, Grzadkowski:2010es}. 
In the SMEFT, the SM fields are classified by their representations in the complete gauge group ${\rm G_{SM}=SU(3)_c\times SU(2)_L\times U(1)_Y}$. 
A perturbative expansion is performed in terms of $1/\Lambda$, where $\Lambda$ denotes the NP scale. 
The leading-order interactions in the SMEFT consist of ${\rm G_{SM}}$ invariant operators up to mass dimension-four (dim-4), i.e. the SM Lagrangian. 
The other is the Higgs effective field theory (HEFT) \cite{Feruglio:1992wf,Bagger:1993zf,Koulovassilopoulos:1993pw,Burgess:1999ha,Grinstein:2007iv,Alonso:2012px,Espriu:2013fia,Buchalla:2013rka,Brivio:2013pma,Alonso:2015fsp,Alonso:2016oah,Buchalla:2017jlu,Alonso:2017tdy,deBlas:2018tjm,Falkowski:2019tft}. 
In the HEFT the fields are classified by their representations in the unbroken gauge group ${\rm SU(3)_c\times U(1)_{\rm em}}$, but the ${\rm SU(2)_L\times U(1)_Y}$ symmetry is nonlinearly realized. 
Since the physical Higgs field $h$ is now a singlet under this nonlinear symmetry, arbitrary functions of $h$ appear in the HEFT Lagrangian. 
In this sense the HEFT already encodes deviations from the SM in the leading order. 
In the HEFT, a perturbative expansion is done according to the number of loops. 

Comparative studies have been actively carried out in recent years on matching a specific ultraviolet (UV) theory onto the SMEFT or the HEFT. 
For instance, Ref.~\cite{Buchalla:2016bse} considers the SM extended with a real singlet scalar and matches it onto the SMEFT and the HEFT respectively. 
The matching of the two-Higgs-doublet model onto both the SMEFT and HEFT has also been performed \cite{Dawson:2023ebe,Dawson:2023oce,Buchalla:2023hqk}. 
In Refs.~\cite{Song:2024kos,Song:2025kjp}, the SM extended with a real triplet scalar is matched onto the HEFT. 
See Refs.~\cite{Alonso:2017tdy,Falkowski:2019tft,Cohen:2020xca} for a more general discussion of theories that cannot be matched onto the SMEFT but must be matched onto the HEFT. 
The classification of the SMEFT and the HEFT in the context of a bottom-up approach is found in Refs.~\cite{Gomez-Ambrosio:2022qsi,Gomez-Ambrosio:2022why,Salas-Bernardez:2022hqv,Delgado:2023ynh,Mahmud:2024iyn}. 
In general, if the NP scale is much higher than the electroweak scale, the new heavy fields are classified by their representations under ${\rm G_{SM}}$. 
By integrating them out, the UV theory is matched to the SMEFT. 
On the other hand, if the NP scale is close to the electroweak scale, the mixing between the SM fields and the new heavy fields cannot be neglected. 
In this case, the theory should be written in terms of the degrees of freedom after the electroweak symmetry breaking (EWSB), and the original electroweak symmetry ${\rm SU(2)_L\times U(1)_Y}$ is nonlinearly realized. 
After bringing them to the mass eigenstates, the new heavy fields are integrated out, resulting in the HEFT.

In this paper, we introduce a new EFT framework that is different from both the SMEFT and the HEFT. 
In our framework, the UV theory is rewritten in terms of the degrees of freedom after the EWSB. 
We do not rewrite it in a nonlinearly realized form of ${\rm SU(2)_L\times U(1)_Y}$, but instead work with the fields in the unitary gauge. 
After bringing the fields to their mass eigenstates, we integrate out the 
physical heavy fields. 
We call thus obtained EFT the broken phase EFT (bEFT). 
In the HEFT, the operators written in nonlinear realizations generate an infinite power of the fields, which complicates the evaluation of the physical observables. 
Our bEFT, on the other hand, does not include an infinite order terms of the fields and is therefore easier to deal with.
The bEFT and the HEFT also differ in their power counting schemes. In the bEFT, a perturbative expansion is made over inverse heavy masses in the broken phase, in contrast to the loop expansion employed in the HEFT.

To illustrate our approach concretely, we consider matching the type-II seesaw model \cite{Konetschny:1977bn,Magg:1980ut,Schechter:1980gr,Cheng:1980qt,Mohapatra:1980yp,Lazarides:1980nt} onto our bEFT. 
In the type-II seesaw model, an ${\rm SU(2)_L}$ triplet scalar is introduced. 
Its vacuum expectation value (VEV) is severely constrained by the $\rho$ parameter measurements. 
The lower bound on the mass of doubly charged scalars is around 400 GeV if the triplet VEV saturates the upper bound on the $\rho$ parameter. 
If the NP scale is close to the electroweak scale, it is expected that the accuracy of the description by the SMEFT deteriorates.
It is important to see whether the SMEFT or the bEFT can describe low-energy physics accurately in such a case. 
Matching the type-II seesaw onto the SMEFT has already been done up to dim-6 operators at the one-loop level in Ref.~\cite{Li:2022ipc}. 
We rewrite its tree-level results in the basis after the EWSB, and call this EFT the broken phase SMEFT (bSMEFT). 
To assess which of the bEFT and bSMEFT can better reproduce the result of the UV theory, we consider two specific processes: 
one is the subprocess at the LHC for the Higgs pair production through vector boson fusion $W^-W^+\to hh$, and the other is the Higgsstrahlung process $e^+e^-\to hZ$ at a future electron-positron collider.
We show that the bEFT reproduces the cross sections of the UV theory more accurately than the bSMEFT over a wide range of parameter space. 
This is because the bEFT deals with the mass eigenstates after symmetry breaking and incorporates the mixing of scalar fields at the leading order.

This paper is organized as follows. 
We introduce the type-II seesaw model in \cref{sec:model}. In \cref{sec:heft}, we describe how to perform its matching onto the bEFT. 
In \cref{sec:smeft}, we rewrite the SMEFT matching results in a basis after the EWSB to get the bSMEFT. 
All operators up to dim-5 and some of dim-6 operators in the bEFT and bSMEFT are shown in \cref{sec:vs}. 
We then evaluate in \cref{sec:num} the two processes mentioned above in the UV theory, bEFT, and bSMEFT respectively and compare their results. 
We conclude in \cref{sec:concl}.

\section{Type-II seesaw model \label{sec:model}}

The type-II seesaw model adds a triplet scalar $\Delta$ to the SM, with the Lagrangian given by 
\begin{align}
\label{eq:model_full-Lagrangian}
{\cal L}_{\tt UV} = {\cal L}_{\tt SM}  + 
\Tr\big[(D_\mu\Delta)^\dag (D^\mu \Delta)\big]
- \left[ \frac{1}{\sqrt{2}} Y_\Delta^{pr}
\overline{L^{\tt C}_p} i \sigma_2 \Delta L_r
 +\hc \right]
 -{\cal V}(H,\Delta),
\end{align}
where the covariant derivative for $\Delta$ is 
\begin{align}
    D_\mu \Delta
    =
    \partial_\mu \Delta
    +i\frac{g}{2}[\sigma^a W^a_\mu, \Delta]
    +ig'B_\mu \Delta  \, ,
\end{align}
with $W_\mu^a~(a=1,2,3)$ and $B_\mu$ denoting the ${\rm SU(2)_L}$ and ${\rm U(1)_Y}$ gauge fields, respectively. The scalar potential is given by 
\begin{align}
    \label{eq:model_potential}
    {\cal V}(H,\Delta)
    =&
    -m_H^2 (H^\dag H)
    +\frac{\lambda}{2}(H^\dag H)^2
    +M_\Delta^2 \Tr(\Delta^\dag \Delta)
    +\frac{\lambda_1}{2}\big[\Tr(\Delta^\dag \Delta)\big]^2
    \nonumber\\
    &
    +\frac{\lambda_2}{2}
     \big( \big[\Tr(\Delta^\dag \Delta)\big]^2
            -\Tr\big[(\Delta^\dag \Delta)^2\big]
     \big)
    +\lambda_4
     (H^\dag H)
     \Tr(\Delta^\dag\Delta)
    \nonumber\\
    &
    +\lambda_5 
     H^\dag
     [\Delta^\dag,\Delta]
     H
    +\bigg( \frac{\Lambda_6}{\sqrt{2}}
            H^T
            i \sigma_2
            \Delta^\dag 
            H
            +\hc
     \bigg) \, ,     
\end{align}
where the notation follows Refs.~\cite{Schmidt:2007nq,BhupalDev:2013xol} and all parameters except for $\Lambda_6$ are real. 
For simplicity, we will also assume $\Lambda_6$ is real. 
The doublet scalar field develops a VEV $v_H$, which through the $\Lambda_6$ coupling induces a VEV $v_\Delta$ for the triplet scalar field: 
\begin{align}
H&=
\frac{1}{\sqrt{2}}
\l( \begin{array}{c}
    \sqrt{2}w^+
    \\
    v_H^{} + h_H + iz_0
    \end{array}
\r) \, ,
\\
\Delta&
=
\frac{\sigma^a}{\sqrt{2}} \Delta^a
=
\frac{1}{\sqrt{2}}
\l( \begin{array}{cc}
    \delta^+ & \sqrt{2}\delta^{++}
    \\
    v_\Delta^{} + \delta^0 + i\eta & -\delta^+
    \end{array}
\r) \, ,
\end{align}
where 
$\Delta^1 = (\delta^{++}+\delta^0)/\sqrt{2}$, $\Delta^2 = i(\delta^{++}-\delta^0)/\sqrt{2}$, and $\Delta^3 = \delta^+$. 
The VEVs are obtained by the minimization conditions of the scalar potential
to be detailed in \cref{app:minimization}, and result in the scalar mixing: 
\begin{subequations}
\begin{align}
&\left( \begin{array}{c}
    G^{\pm}
    \\
    H^{\pm}
    \end{array}     
\right) 
=
\left( \begin{array}{cc}
    \cos\beta' & \sin\beta'
    \\
    -\sin\beta' & \cos\beta'
    \end{array}     
\right) 
\left( \begin{array}{c}
    w^\pm
    \\
    \delta^\pm
    \end{array}     
\right) \, ,
\\
&\left( \begin{array}{c}
    h
    \\
    H_0
    \end{array}     
\right) 
=
\left( \begin{array}{cc}
    \cos\alpha & \sin\alpha
    \\
    -\sin\alpha & \cos\alpha
    \end{array}     
\right) 
\left( \begin{array}{c}
    h_H
    \\
    \delta^0
    \end{array}     
\right) \, ,
\\
&\left( \begin{array}{c}
    G_0
    \\
    A_0
    \end{array}     
\right) 
=
\left( \begin{array}{cc}
    \cos\beta & \sin\beta
    \\
    -\sin\beta & \cos\beta
    \end{array}     
\right) 
\left( \begin{array}{c}
    z^0
    \\
    \eta
    \end{array}     
\right) \, ,
\end{align}
\end{subequations}
where the mixing angles are given by 
\begin{subequations}
\begin{align}
   \label{eq:model_tanBetaPrime} 
    &\tan\beta'
    =
    \frac{\sqrt{2}v_\Delta}{ v_H^{} } \, ,
    \\
    \label{eq:model_tanBeta}
    &\tan\beta
    =
    \frac{2v_\Delta}{ v_H^{} }
    =
    \sqrt{2}\tan\beta' \, ,
    \\
    \label{eq:model_tan2Alpha}
    &\tan2\alpha
    =
    \frac{4v_\Delta}{ v_H^{} }
    \frac{M_\Delta^2 + \frac{1}{2}\lambda_1v_\Delta^2}{M_\Delta^2+(\lambda_- -\lambda) v_H^2 +\frac{3}{2}\lambda_1v_\Delta^2} \, ,
\end{align}
\end{subequations}
with 
\begin{align}
\lambda_{-} = \frac{1}{2}(\lambda_4 - \lambda_5) \, .
\end{align}

The triplet VEV breaks the custodial symmetry, and is severely constrained by the precision $\rho$ parameter. 
The latter is expressed in terms of the $W$ boson mass $M_W$, the $Z$ boson mass $M_Z$, and the Weinberg angle $\theta_W$:
\begin{align}
    \rho
    =\frac{M_W^2}{M_Z^2\cos^2\theta_W}
    =\frac{1}{1+2\epsilon ^2}\, ,
\end{align}
where 
\begin{align}
\epsilon =\frac{v_\Delta}{v_{\rm ew}},
\end{align}
and $v_{\rm ew}=\sqrt{v_H^2+2v_\Delta^2} \approx 246\,{\rm GeV}$. 
Allowing a $3\sigma$ deviation from the measured value $\rho_{\rm exp} = 1.00038 \pm 0.00020$ \cite{ParticleDataGroup:2022pth}, 
we obtain the upper bound $\epsilon \leq 0.0105$. 
In the following sections, we will work up to order $\epsilon^2$. 
We list here some interactions in the unitary gauge, including the ones relevant to later evaluation of the processes $W^-W^+ \to h h$ and $e^+e^-\to hZ$ in \cref{sec:num}:
\begin{align}
\mathcal{L}_{\tt UV}
\supset&\,
C_{h^3}^{\tt UV} h^3
+C_{hW^2}^{\tt UV} h W^-_\mu W^{+\mu} 
+C_{h^2W^2}^{\tt UV} h^2 W^-_\mu W^{+\mu}
+C_{hZ^2}^{\tt UV} h Z_\mu Z^\mu
+C_{h^2Z^2}^{\tt UV} h^2 Z_\mu Z^\mu 
\nn\\
&\,
+C_{h^2H_0}^{\tt UV} h^2H_0
+C_{H_0W^2}^{\tt UV} H_0 W_\mu^-W^{+\mu}
+C_{H_0Z^2}^{\tt UV} H_0Z_\mu Z^\mu
\nn\\
&\,
+\big(
C_{hHW}^{\tt UV} (H^+ i \overset{\leftrightarrow}{D}_\mu h)W^{-\mu}
+\hc\big)
+C_{hA_0Z}^{\tt UV} (A_0\overset{\leftrightarrow}{\partial}_\mu h) Z^\mu
\, ,
\end{align} 
where $\Phi_1 \overset{\leftrightarrow}{D}_\mu \Phi_2\equiv  \Phi_1 D_\mu \Phi_2 - (D_\mu \Phi_1) \Phi_2 $, and the coefficients are 
\begin{subequations}
\begin{align}
\label{eq:model_h3}
&C_{h^3}^{\tt UV}  
=-\frac{\lambda}{2}v_{\rm ew}^{}
    +\epsilon^2 v_{\rm ew}
        \l[  \frac{(6\kappa^3-2\kappa^2+\kappa-1)}{2r_\Delta^2\kappa}
            -\frac{\lambda_-
            (2\kappa^2-1)}{2}  
        \r] \, ,
\\
\label{eq:model_hww}
&C_{hW^2}^{\tt UV}
=\frac{g^2v_{\rm ew}^{}}{2} 
     [ 1+\epsilon^2  ( -2\kappa^2+4\kappa -1 ) ] \, ,        
\\
&C_{h^2W^2}^{\tt UV}
=\frac{g^2}{4} [ 1 + 4\epsilon^2 \kappa^2 ] \, ,
\\
&C_{hZ^2}^{\tt UV}
=\frac{g^2 v_{\rm ew}^{}}{4c_W^2} 
    [ 1+\epsilon^2  ( -2\kappa^2 + 8\kappa -1  ) ] \, ,
\\
&C_{h^2Z^2}^{\tt UV}
=\frac{g^2}{8c_W^2} [ 1 + 12\epsilon^2 \kappa^2 ] \, , 
\\
&C_{h^2H_0}^{\tt UV}
=\epsilon v_{\rm ew}^{}
\frac{3\kappa-2 -\lambda_-
    \kappa r_\Delta^2}{r_\Delta^2} \, ,
\\
\label{eq:model_H0ww}
&C_{H_0W^2}^{\tt UV}
=-g^2\epsilon(\kappa-1)v_{\rm ew}^{} \, ,
\\
&C_{H_0Z^2}^{\tt UV}
=
-\epsilon v_{\rm ew}^{}\frac{g^2(\kappa-2)}{2c_W^2} \, ,
\\
&C_{hHW}^{\tt UV}
=\epsilon\frac{g(2\kappa-1)}{\sqrt{2}} \, ,
\\
\label{eq:model_hA0Z}
&C_{hA_0Z}^{\tt UV}
=\epsilon\frac{g(2\kappa-1)}{c_W^{}} \, ,
\end{align}
\end{subequations}
where 
\begin{align}
r_\Delta = \frac{v_{\rm ew}}{M_\Delta} \, , \quad 
\kappa =\frac{1}{1 + r_\Delta^2(\lambda_- -\lambda)} \, .
\end{align}
One notices that the new contributions from the triplet scalar are suppressed by ${\cal O}(\epsilon)$ or ${\cal O}(\epsilon^2)$ relative to the SM parts.

\section{Matching onto bEFT \label{sec:heft}}

In this section, we perform the matching of the type-II seesaw model onto the bEFT at tree level. 
As we emphasized in \cref{sec:intro}, the bEFT introduced in this work differs from the HEFT in several ways. 
The HEFT employs a perturbative expansion in the number of loops. 
Our bEFT, on the other hand, employs double expansions in terms of both $\epsilon$ and the inverse of heavy fields' masses. 
In addition, the HEFT introduces the would-be Goldstone fields in a nonlinear realization, while our bEFT assumes the unitary gauge and removes the would-be Goldstone fields from the very start. 
We take the new scalars $H_0$, $A_0$, $H^\pm$, and $H^{\pm\pm}$ as heavy states and integrate them out to perform matching onto the bEFT.

Before going into the details, we introduce an auxilliary $\mathbb{Z}_2$ parity that can help bookkeep the power of expansion on the matching. 
Under the $\mathbb{Z}_2$ the triplet scalar and the couplings $Y_\Delta,~\Lambda_6$ flip sign, $(\Delta, Y_\Delta, \Lambda_6) \to - (\Delta, Y_\Delta, \Lambda_6)$, while the SM fields and other couplings keep intact. 
This symmetry is preserved in the broken phase since the triplet VEV being proportional to $\Lambda_6$ is also odd under the $\mathbb{Z}_2$, i.e., $v_\Delta \to - v_\Delta$ or $\epsilon \to -\epsilon$. 
The solutions of the equations of motion for the heavy scalar fields $\Phi \in  \{H_0,A_0,H^\pm,H^{\pm\pm}\}$ are expressed in terms of the SM fields in the mass eigenstates. 
The odd parity of the heavy fields implies the solutions in the form: 
\begin{align}
\label{eq:heft_eom-sol}
\Phi_c[\phi_{\tt SM}]
=
\sum_{m=0}^\infty
\sum_{n=0}^{2m+1}
\epsilon^n Y_\Delta^{2m+1-n} 
\mathcal{O}^{(m,n)}[\phi_{\tt SM}] \, ,
\end{align}
where $\mathcal{O}^{(m,n)}[\phi_{\tt SM}]$ is an operator composed only of the SM fields $\phi_{\tt SM}$. 
Our goal is to match the type-II seesaw model onto the bEFT up to dim-6 and order $\epsilon^2$. 
In light of the fact that $\mathcal{O}^{(m,n)}[\phi_{\tt SM}]$ contains lepton bilinears $(\overline{l_1^{\tt C}}l_2)^{2m+1-n}$ with $l=\ell, \nu$, we can easily check that the following three terms in the scalar potential, 
\begin{align}
\label{eq:heft_Lir}
&\mathcal{V}^{\tt bEFT}[h,\Phi_c]
\supset
\tilde{\lambda}_1 \epsilon  \Phi_c^3
+\tilde{\lambda}_2 \epsilon h \Phi_c^3 
+ \tilde{\lambda}_3 \Phi_c^4  \, ,
\end{align}
only generate operators of dim-$n~(n\geq 7$) and/or of order $\epsilon^n~(n\geq 3)$, and are therefore irrelevant.
Here $\tilde{\lambda}_i~(i=1,2,3)$ is given by a function of parameters in the original Lagrangian.
The remaining interactions relevant to our analysis are given by 
\begin{align}
\label{eq:heft_V-relevant}
\mathcal{V}^{\tt UV}[h,\Phi]
&\supset
\frac{m_{h}^2}{2} h^2
+\frac{m_{H_0}^2}{2} H_0^2
+\frac{m_{A_0}^2}{2} A_0^2
+m_{H^\pm}^2 H^-H^+
+m_{H^{\pm\pm}}H^{--}H^{++}
\nn\\
&\quad
+{1\over 2} (2v_{\rm ew}^{}h + h^2)
\big[
\lambda_- H_0^2
+  \lambda_- A_0^2
+  (2\lambda_- + \lambda_5)  H^-H^+
+ 2 (\lambda_- +\lambda_5)  H^{--}H^{++}
\big]
\nn\\
&\quad
+ \epsilon v_{\rm ew}^{}
  \l[  \frac{M_\Delta^2(2-3\kappa)}{v_{\rm ew}^2} + \lambda_-\kappa  \r] h^2 H_0 
+ \epsilon
  \l[  \frac{M_\Delta^2(1-\kappa)}{v_{\rm ew}^2} + \lambda_-\kappa  \r] h^3 H_0 \,, 
\end{align}
and the kinetic energy terms of heavy fields. 
We can now integrate out heavy scalar fields $H_0$, $A_0$, $H^\pm$, and $H^{\pm\pm}$ to obtain the bEFT up to the dim-6 level. 
This may be accomplished with the help of the Mathematica package {\tt Matchete}~\cite{Fuentes-Martin:2022jrf}. The resulting effective operators and their Wilson coefficients (WCs) are listed in \cref{sec:vs}.

\section{Matching onto bSMEFT\label{sec:smeft}}

In this section, we consider matching the type-II seesaw model onto the SMEFT and rewrite it in terms of the fields in the broken phase to get the bSMEFT. 
We emphasize here that the bEFT in the previous section is obtained by integrating out heavy fields after the EWSB, 
whereas the bSMEFT is obtained by integrating out heavy fields before the EWSB. 
The tree level matching result onto the SMEFT can be found in Ref.~\cite{Li:2022ipc}, which is given by
\begin{align}
\label{eq:smeft_Leff}
\mathcal{L}_{\tt SMEFT}^{\rm dim6}
=&
\mathcal{L}_{\tt SM}
+\frac{\Lambda_6^2}{2M_\Delta^2} 
    \Big(  1 - \frac{2m_H^2}{M_\Delta^2}  \Big)
    (H^\dag H)^2
+
\big( C^{(5)}_{pr}\mathcal{O}^{(5)}_{pr} + \hc \big)         
\nn\\
&\,
+C_{\ell\ell}^{prst} \mathcal{O}_{\ell\ell}^{prst}
+C_{H} \mathcal{O}_{H}
+C_{H\square} \mathcal{O}_{H\square}
+C_{H D} \mathcal{O}_{H D}
\nn\\
&\,
+
\big(  C_{eH}^{pr} \mathcal{O}_{eH}^{pr}
         +C_{uH}^{pr} \mathcal{O}_{uH}^{pr}
         +C_{dH}^{pr} \mathcal{O}_{dH}^{pr}
         +\hc
\big) \, ,      
\end{align}
where the dim-5 Weinberg operator is $\mathcal{O}_{pr}^{(5)} = \bar{L}^p \tilde{H} \tilde{H}^T L^{r \tt C}$ and the remaining dim-6 operators adopt the standard Warsaw basis notation \cite{Grzadkowski:2010es}. 
The matched WCs are given by
\begin{subequations}
\begin{align}
&C^{(5)}_{pr} 
= -\frac{\Lambda_6 Y_\Delta^{*pr} }{2M_\Delta^2} \, ,
\quad
C_{\ell\ell}^{prst} 
= \frac{Y_\Delta^{*ps} Y_\Delta^{rt}}{4M_\Delta^2}  \, , 
\nn
\\
&C_H 
= (4\lambda-\lambda_-) \frac{\Lambda_6^2}{M_\Delta^4}
-\frac{\Lambda_6^4}{M_\Delta^6} \, ,
\nn
\\
&C_{H\square} = \frac{\Lambda_6^2}{2M_\Delta^4} \, ,
\quad
C_{HD} = \frac{\Lambda_6^2}{M_\Delta^4} \, ,
\nn
\\
&C_{eH}^{pr} = \frac{\Lambda_6^2  Y_l^{pr}}{2M_\Delta^4} \, ,
\quad
C_{uH}^{pr} = \frac{\Lambda_6^2 Y_{\rm u}^{pr}}{2M_\Delta^4}  \, ,
\quad
C_{dH}^{pr} = \frac{\Lambda_6^2 Y_{\rm d}^{pr}}{2M_\Delta^4}  \, ,
\end{align}
\end{subequations}
where the superscripts $p,r,s,t$ denote fermion generations. 
Note that $\Lambda_6$ is considered as the same order as the bare mass $M_\Delta$ in the SMEFT power counting. 
$Y_l^{pr}$, $Y_{\rm u}^{pr}$, and $Y_{\rm d}^{pr}$ are the SM Yukawa couplings for the charged leptons, up-type quarks, and down-type quarks, respectively. 
In the following part, we will rewrite the above SMEFT interactions in terms of the broken phase fields to get the bSMEFT. 

\subsection*{Shift of VEV}

The scalar potential is modified by the inclusion of the dim-6 operator $\mathcal{O}_H$: 
\begin{align}
{\cal V}_{\tt SMEFT}^{\rm dim6}
=
-m_H^2 H^\dag H
+\frac{\lambda_{\rm eff}}{2}(H^\dag H)^2
-C_H (H^\dag H)^3 \, ,
\end{align}
with
\begin{align}
\label{eq:smeft_lambda_eff}
\lambda_{\rm eff}
=
\lambda  -\frac{\Lambda_6^2}{M_\Delta^2} 
    \l(  1 - \frac{2m_H^2}{M_\Delta^2}  \r) \, .
\end{align}
The dimensionful parameter $\Lambda_6$ can be expressed in terms of $v_H^{}$ and $v_\Delta^{}$ by \cref{eq:model_minimization-2}: 
\begin{align}
\label{eq:smeft_Lambda6}
\Lambda_6
=
\frac{2M_\Delta^2v_\Delta + \lambda_1 v_\Delta^3 + 2\lambda_- v_\Delta v_H^2}{v_H^2} \, . 
\end{align}
The solution for the new minimum is given by
\begin{align}
\label{eq:smeft_PhiSqVEV}
\langle H^\dag H \rangle
=
\frac{\lambda_{\rm eff} - \sqrt{\lambda_{\rm eff}^2 - 6 C_H \lambda_{\rm eff} v_H^2}}{6C_H} \, .
\end{align}
We expand the above solution to the first order of $C_H$ and define the new vacuum as 
\begin{align}
\langle H^\dag H \rangle
=
\frac{v_H^2}{2}
  \l(  1 
       + \frac{3C_H v_H^2}{2\lambda_{\rm eff}}
  \r)
\equiv
\frac{\tilde{v}_{\rm ew}^2}{2} \, .
\end{align}
Note that $\tilde{v}_{\rm ew}^{}$ can be expressed in terms of the potential parameters as 
\begin{align}
    \tilde{v}_{\rm ew}^2
    =
    v_H^2
    \l[ 1 
        + \frac{6 (14\lambda - \lambda_-)\big(M_\Delta^2 +  \lambda_- v_H^2 \big)^2}{\lambda M_\Delta^4} \frac{v_\Delta^2}{v_H^2}
        + \mathcal{O}(v_\Delta^3/v_H^3)
    \r] \, .
\end{align}
We emphasize that, as the notation suggests, the above expression for $\tilde{v}_{\rm ew}^{}$ is different from that for $v_{\rm ew}^{}$ introduced in \cref{eq:model_vT} for the UV theory, although they are numerically identical: 
$v_{\rm ew} = \tilde{v}_{\rm ew} = 246\,{\rm GeV}$. 
In the following, as we did for the UV theory in \cref{sec:model}, we expand the bSMEFT in terms of the parameter
\begin{align}
\tilde{\epsilon} =  \frac{v_\Delta}{\tilde{v}_{\rm ew}}.
\end{align}

\subsection*{Higgs-gauge interactions}

The interactions between the Higgs and gauge bosons are modified by the dim-6 operators $\mathcal{O}_{H \square}$ and $\mathcal{O}_{H D}$ which will change the normalization of the Higgs field. 
Redefining the Higgs field to make it canonically normalized will modify all interactions involving the Higgs field. 
The relevant terms are 
\begin{align}
\label{eq:smeft_LkinNLO}
\mathcal{L}_{\tt SMEFT}^{H,{\rm kin}}
=(D_\mu H)^\dag D^\mu H
+ C_{H \square} \mathcal{O}_{H \square}
+ C_{H D} \mathcal{O}_{H D} \, ,
\end{align}
yielding the derivative terms of the Higgs field $h_H$: 
\begin{align}
\label{eq:smeft_LkinNLO-2}
&\mathcal{L}_{\tt SMEFT}^{H,{\rm kin}}
\supset
\frac{1}{2}
    \l[ 1 
        - 2c_{\rm kin}^{}   
        \l( 1+\frac{h_H}{\tilde{v}_{\rm ew}^{}} \r)^2 
    \r] \partial_\mu h_H \partial^\mu h_H \, ,
\end{align}
where 
\begin{align}
\label{eq:smeft_ckin}
c_{\rm kin}
=
\tilde{v}_{\rm ew}^2 C_{H \square}
-\frac{1}{4}\tilde{v}_{\rm ew}^2 C_{H D}
=
\tilde{\epsilon}^2
\l( 1 + \lambda_- \tilde{r}_\Delta^2 \r)^2 \,, \quad  
\tilde{r}_\Delta \equiv \frac{\tilde v_{\rm ew}}{M_\Delta}.
\end{align}
The Higgs field is made canonically normalized by rescaling 
\begin{align}
\label{eq:smeft_htilde}
\tilde{h}=h_H\sqrt{1-2c_{\rm kin}}
\approx h_H(1-c_{\rm kin})\, .
\end{align}
By substituting \cref{eq:smeft_htilde}
into 
\cref{eq:smeft_LkinNLO}, we get modified Higgs-gauge interactions.

\subsection*{Higgs self-interactions}

The Higgs self-interactions are modified by the operator $\mathcal{O}_{H}$, as well as by the Higgs field redefinition through the $\mathcal{O}_{H \square}$ and $\mathcal{O}_{HD}$ operators. 
The final form of the scalar potential in the broken phase is given by
\begin{align}
\label{smeft_scalar-potential}
\mathcal{L}_{\tt SMEFT}^{\rm dim6}
\supset&\,
-\frac{\lambda_{\rm eff}}{2} 
\tilde{v}_{\rm ew}^2
\l(  1 
     - \frac{3C_H v_H^2}{\lambda_{\rm eff}}
     +2c_{\rm kin}
\r) \tilde{h}^2
-
\frac{\lambda_{\rm eff}}{2}
\tilde{v}_{\rm ew}^{}     
\l(  1
     -\frac{5C_H \tilde{v}_{\rm ew}^2}{\lambda_{\rm eff}}
     +3 c_{\rm kin}
\r) \tilde{h}^3    
\nn\\
&\,
-
\frac{\lambda_{\rm eff}}{8}
\l(  1
     -\frac{15 C_H \tilde{v}_{\rm ew}^2}{\lambda_{\rm eff}}  
     +4c_{\rm kin}
\r) \tilde{h}^4
+
\frac{3}{4} C_H v_H^{} \tilde{h}^5
+
\frac{1}{8} C_H \tilde{h}^6 \, .    
\end{align}

\subsection*{Yukawa interactions and neutrino mass}

The Yukawa interactions are modified by the operator $\mathcal{O}_{\psi H}^{pr}~(\psi = {u,d,e})$. 
Including the Higgs field redefinition, they become 
\begin{align}
\mathcal{L}_{\tt SMEFT}^{\rm dim6}
\supset&\,
-\frac{\tilde{v}_{\rm ew}^{}}{\sqrt{2}}
\Big( 1+(1+c_{\rm kin})\frac{\tilde{h}}{\tilde{v}_{\rm ew}^{}} \Big)
\sum_{\psi = u,d,e}\Big[ Y_\psi^{pr}
    -\frac{C_{\psi H}^{pr}\tilde{v}_{\rm ew}^2}{2M_\Delta^2}
    \Big( 1 + \frac{\tilde{h}}{\tilde{v}_{\rm ew}^{}} \Big)^2 
\Big]
\bar{\psi}_L^p\psi_R^r   
+
\hc \, .
\end{align}
The neutrino mass and neutrino Yukawa interactions are induced by the dim-5 operator $O^{(5)}_{pr}$, which become in the broken phase 
\begin{align}
C_{pr}^{(5)} \mathcal{O}_{pr}^{(5)} + {\tt h.c.}
=
-\frac{\tilde{v}_{\rm ew}^2}{2} 
    C_{pr}^{(5)}
    \Big( 1+\frac{\tilde{h}}{\tilde{v}_{\rm ew}} \Big)^2~
    \overline{\nu_{L,p}^{\tt C}} 
    \nu_{L,r}
+\hc \, .
\end{align}

\section{List of Wilson coefficients for bEFT and bSMEFT operators \label{sec:vs}}

In this section we list the WCs of the effective operators in the bEFT and bSMEFT. 
In all tables the second column lists the WCs for the operators in the bEFT up to order $\epsilon^2$, while the third column lists those in the bSMEFT up to order $\tilde{\epsilon}^2$ and $\tilde{r}_\Delta^2$. 
The field $h$ in the third column corresponds to the canonically normalized $\tilde{h}$ in Sec.~\ref{sec:smeft}.
Note that the Hermitian conjugate of a non-Hermitian operator is not listed.

\begin{table}[t]
\begin{center}
\resizebox{\linewidth}{!}{
\renewcommand\arraystretch{1.6}
\begin{tabular}{|c||c|c|}
\hline
\rowcolor{lightgray} \multicolumn{3}{|c|}{\bf dim-3} \\
\hline
Operator
& WC in bEFT 
& WC in bSMEFT \\
\hline
\multicolumn{3}{|c|}{$\Delta L =0$} 
\\
\hline
$h^3$
& $\displaystyle
    -\frac{\lambda}{2}v_{\rm ew}^{}
    +\epsilon^2 v_{\rm ew}^{}
        \l[ \frac{(1-2\kappa^2)\lambda_-}{2} - \frac{1-\kappa+2\kappa^2 - 6\kappa^3}{2 \kappa r_\Delta^2 }
        \r]$
& $\displaystyle
    -\frac{\lambda}{2} \tilde{v}_{\rm ew}^{}
    +\tilde{\epsilon}^2 \tilde{v}_{\rm ew}^{}
        \bigg[ 
            \frac{2}{\tilde{r}_\Delta^2}
            +\frac{73}{2}\lambda
            -6 \lambda_-
            +
         \lambda_{-}(73\lambda - 18 \lambda_-)\tilde{r}_\Delta^2
        \bigg]$
\\ 
$hW^-_\mu W^{+\mu}$
& $\displaystyle 
    \frac{g^2v_{\rm ew}^{}}{2} 
    \Big[ 1 - \epsilon^2  ( 1 - 4 \kappa + 2\kappa^2 ) \Big]$ 
& $\displaystyle  
    \frac{g^2 \tilde{v}_{\rm ew}^{}}{2}  
    \Big[ 
        1 
        +\tilde{\epsilon}^2
        \big( 1 + 2 \lambda_- \tilde{r}_\Delta^2  \big)
    \Big]$ \\
$h Z_\mu Z^\mu$
& $\displaystyle 
    \frac{g^2 v_{\rm ew}^{}}{4c_W^2} 
        \Big[ 1 - \epsilon^2 ( 1 - 8 \kappa + 2\kappa^2 )\Big]$ 
& $\displaystyle 
    \frac{g^2 \tilde{v}_{\rm ew}}{4c_W^2} 
        \Big[ 
            1
            +5\tilde{\epsilon}^2
            \big( 1 + 2 \lambda_- \tilde{r}_\Delta^2   \big)
        \Big]$ \\
\hline
\multicolumn{3}{|c|}{$\Delta L =2$} \\
\hline
$\overline{\nu_{L,p}^{\tt C}} \nu_{L,r}$
& $\displaystyle 
    -\epsilon\frac{v_{\rm ew}^{}}{2} Y_\Delta^{pr} $ 
& $\displaystyle 
    -\tilde{\epsilon} \frac{\tilde{v}_{\rm ew}^{}}{2} Y_\Delta^{pr} 
        \l[ 1 + \lambda_- \tilde{r}_\Delta^2  \r]$ \\
\hline 
\end{tabular}
}
\caption{
WCs for the dim-3 operators in the bEFT and bSMEFT.
}
\label{table:heft-vs-smeft_3d}
\end{center}
\end{table}

\begin{table}[t]
\begin{center}
\resizebox{\linewidth}{!}{
\renewcommand\arraystretch{1.6}
\begin{tabular}{|c||c|c|}
\hline
\rowcolor{lightgray} \multicolumn{3}{|c|}{\bf dim-4} \\
\hline
Operator
& WC in bEFT 
& WC in bSMEFT \\
\hline
\multicolumn{3}{|c|}{$\Delta L =0$} \\
\hline
$h^4$
& $\displaystyle
    -\frac{\lambda}{8}
    - \epsilon^2 
    \Bigg[  \frac{(1-\kappa)\kappa }{r_\Delta^2}
        +\kappa^2 \lambda_-
        + \frac{ ( 2 - 3\kappa + \kappa \,\lambda_- r_\Delta^2  )^2}{2 (1 + \lambda_- r_\Delta^2 ) r_\Delta^2 } 
    \Bigg]$ 
& $\displaystyle
    -\frac{\lambda}{8}
    +\tilde{\epsilon}^2
    \bigg[  \frac{1}{2\tilde{r}_\Delta^2}
        +29\lambda 
        -\frac{13}{2}\lambda_-
        + \frac{29}{2}\lambda_- (4\lambda - \lambda_-) \tilde{r}_\Delta^2   
    \bigg]$ \\ 
$h^2W^-_\mu W^{+\mu}$ 
& $\displaystyle 
    \frac{g^2}{4}  
        \l[ 1 
            - \epsilon^2
            \frac{4 (1-2\kappa) \l(2 -\kappa + \kappa\, \lambda_- r_\Delta^2  \r)
            }{1 + \lambda_- r_\Delta^2 } 
        \r]$ 
& $\displaystyle 
    \frac{g^2}{4} 
        \big[ 1 + 2\tilde{\epsilon}^2 \big( 1 + 2 \tilde{r}_\Delta^2 \lambda_- \big) \big]$ \\
$h^2 Z_\mu Z^\mu$
& $\displaystyle 
    \frac{g^2}{8c_W^2}
        \l[ 1 
            -\epsilon^2
            \frac{8(1-2\kappa) (2 + \kappa\, \lambda_- r_\Delta^2 )
            }{1 + \lambda_-  r_\Delta^2 }  
        \r]$ 
& $\displaystyle 
    \frac{g^2}{8c_W^2} 
        \big[ 1 +14\tilde{\epsilon}^2 \big( 1 + 2 \tilde{r}_\Delta^2 \lambda_- \big) \big]$ \\     
$W^-_\mu W^{+\mu} W^-_{\nu}W^{+\nu}$
& $\displaystyle 
    -\frac{g^4}{2}
        \l[ 1 
            -\epsilon^2
            \frac{ (1-\kappa)^2 r_\Delta^2  }{1 + \lambda_-  r_\Delta^2} 
        \r]$ 
& $\displaystyle 
    -\frac{g^4}{2}$ \\
$W^-_\mu W^+_\nu W^{-\mu} W^{+\nu}$
& $\displaystyle 
    \frac{g^4}{2}
        \l[ 1 
            +\epsilon^2
            \frac{ r_\Delta^2}{1 + \lambda_- r_\Delta^2 } 
        \r]$ 
& $\displaystyle 
    \frac{g^4}{2}$ \\
$W^-_\mu W^{+\mu} Z_{\nu} Z^{\nu}$
& $\displaystyle 
    -g^4c_W^2
        \l[ 1 
            -\epsilon^2
            \frac{(2 - \kappa)(1-\kappa) r_\Delta^2
            }{2 c_W^4 ( 1 + \lambda_- r_\Delta^2 )} 
        \r]$ 
& $\displaystyle -g^4c_W^2$ \\
$W^-_\mu W^+_{\nu} Z^{\mu} Z^{\nu}$
& $\displaystyle 
    g^4c_W^2
        \l[ 1 
            +\epsilon^2
            \frac{r_\Delta^2}{c_W^4 \big( 2 + (2\lambda_{-}+\lambda_5) r_\Delta^2 \big)}
        \r]$ 
& $\displaystyle  g^4c_W^2$ \\
$Z_\mu Z^{\mu} Z_\nu Z^{\nu}$ 
& $\displaystyle 
    \epsilon^2  \frac{g^4 (2- \kappa)^2 r_\Delta^2
    }{ 8 c_W^4 ( 1 + \lambda_-  r_\Delta^2 ) }   $ 
& 0 \\
$h\bar{\psi}_{L,p}\psi_{R,r} (\psi = u,d,e)$
& $\displaystyle 
    -\frac{1}{\sqrt{2}} Y_\psi^{pr}$
& 
$\displaystyle 
    -\frac{1}{\sqrt{2}} Y_\psi^{pr}
    \big[ 1-2\tilde{\epsilon}^2 \big( 1 + 2 \lambda_- \tilde{r}_\Delta^2 \big) \big]$ \\    
\hline
\multicolumn{3}{|c|}{$\Delta L =2$} \\
\hline
$h\l( \overline{\nu_{L,p}^{\tt C}} 
    \nu_{L,r} \r)$ 
& $\displaystyle 
    -\tilde{\epsilon} Y_\Delta^{pr} (1 + \lambda_- r_\Delta^2 )$ 
& $\displaystyle 
    -\tilde{\epsilon} Y_\Delta^{pr} ( 1 + \lambda_- \tilde{r}_\Delta^2 )$ \\    
\hline
\end{tabular}
}
\caption{ WCs for the dim-4 operators in the bEFT and bSMEFT.
}
\label{table:heft-vs-smeft_4d}
\end{center}
\end{table}

In \cref{table:heft-vs-smeft_3d} and \cref{table:heft-vs-smeft_4d}, we list the coefficients of the lepton-number-conserving ($\Delta L=0$) and -violating ($\Delta L=2$) dim-3 and dim-4 operators, respectively. 
Notice that we have expressed the matched WCs in both EFTs in terms of the common set of parameters in order to facilitate numerical comparison. The heavy masses in WCs of the bEFT have been replaced by the chosen parameters that make them look a little bit complicated.
As can be seen from \cref{table:heft-vs-smeft_3d}, all the coefficients of dim-3 operators in the bEFT coincide with those in the UV.
From these tables, the deviations from the SM in the Higgs-gauge and pure-gauge interactions are of order $\epsilon^2$ ($\tilde{\epsilon}^2$) in the bEFT (bSMEFT). 
This can be understood from the $\mathbb{Z}_2$ symmetry introduced in \cref{sec:heft}. 
The expressions for the neutrino mass differ between the bSMEFT and the bEFT. 
This is because in the bSMEFT the neutrino mass term arises through the Weinberg operator, whereas in the bEFT the neutrino mass appears already before integrating out heavy fields through the mixing among scalar fields. 
Note that the $W^4$, $W^2Z^2$, and $Z^4$ interactions in the bSMEFT coincide with those in the SM as can be seen in \cref{table:heft-vs-smeft_4d}.
This is because the pure gauge operators without derivatives are generated only through the product of the $D_\mu H$, so that the corrections to these interactions start to arise only from the dim-8  operators.

In \cref{table:heft-vs-smeft_5d} we list the WCs of both lepton-number-conserving and lepton-number-violating dim-5 operators in the bEFT and bSMEFT. 
Similar comments to the above quartic $W,~Z$ operators apply to the dim-5 pure bosonic operators as well. 
Note that the bSMEFT includes $h^3Z^2$ but not $h^3W^2$ interactions, reflecting custodial symmetry breaking by $\mathcal{O}_{HD}$. 
And finally in \cref{table:heft-vs-smeft_6d}, we show the WCs of the dim-6 operators in the bEFT and the bSMEFT that are relevant to the $W^-W^+\to hh$, $ZZ\to hh$, and $e^+e^-\to hZ$ scattering processes, two of which will be computed in the next section.

\begin{table}[t]
\begin{center}
\resizebox{\linewidth}{!}{
\renewcommand\arraystretch{1.6}
\begin{tabular}{|c||c|c|}
\hline
\rowcolor{lightgray} \multicolumn{3}{|c|}{\bf dim-5} \\
\hline
Operator
& WC in bEFT 
& WC in bSMEFT \\ 
\hline
\multicolumn{3}{|c|}{$\Delta L =0$} \\
\hline
$h^5$
& 
    $\displaystyle
    \frac{\epsilon^2  \big[ (1-\kappa)(2-3\kappa)r_\Delta^{-2} -2 (1 -5\kappa+ 5\kappa^2 )\lambda_{-}
        -  (1-3\kappa)\kappa\,\lambda_{-}^2 r_\Delta^2  \big ] }{ v_{\rm ew} (1 + \lambda_{-} r_\Delta^2)^2} $
&  $\displaystyle 
    \frac{3 \tilde{\epsilon}^2}{\tilde{v}_{\rm ew}^{}}
    ( 4 \lambda-\lambda_{-}) (1  +2  \lambda_{-}\tilde{r}_\Delta^2 ) $ \\    
$h^3 W^-_\mu W^{+\mu}$
& 
    $\displaystyle
    \frac{\epsilon^2 g^2\big[ - 1+ 2\kappa^2 + (1-\kappa)(3-8\kappa) \lambda_{-} r_\Delta^2
    + (1-2\kappa)\,\kappa\lambda_{-}^2 r_\Delta^4  \big] }{ v_{\rm ew} ( 1 + \lambda_{-}r_\Delta^2 )^2}  $
& $0$ \\    
$h W^-_\mu W^{+\mu} W^-_\nu W^{+\nu}$
& $\displaystyle
        \frac{ \epsilon^2 g^4(1-\kappa) 
        \big[ \kappa \, r_\Delta^2 - (1-2\kappa) \lambda_{-}r_\Delta^4 \big]
        }{ v_{\rm ew} (1 + \lambda_{-} r_\Delta^2 )^2}$ 
& 0 \\
$h W^-_\mu W^+_\nu W^{-\mu} W^{+\nu}$
& $\displaystyle 
    \frac{
        \epsilon^2 g^4 
        \big[  2 \kappa\, r_\Delta^2 - (1-2\kappa)\lambda_{-} r_\Delta^4  \big]  
        }{
        v_{\rm ew}(1 +(\lambda_{-}+\lambda_5) r_\Delta^2)^2} $ 
& 0 \\
$h^3 Z_\mu Z^\mu$
& $\displaystyle
    \frac{\epsilon^2g^2\big[- 2 - 3 \kappa + 8 \kappa^2     +3  (2-7\kappa+4\kappa^2) \lambda_{-} r_\Delta^2
    + 2 (1-2\kappa )\kappa\,\lambda_{-}^2 r_\Delta^4 \big] }{ 2 v_{\rm ew} c_W^2 ( 1 + \lambda_{-} r_\Delta^2 )^2}  $
& $\displaystyle 
    \frac{\tilde{\epsilon}^2g^2 (1+2\lambda_{-} \tilde{r}_\Delta^2)}{
    \tilde{v}_{\rm ew}c_W^2}$ \\    
$h Z_\mu Z^\mu Z_\nu Z^\nu$ 
& $\displaystyle 
   \frac{
        \epsilon^2g^4(2-\kappa)
        \big[  3 \kappa\,r_\Delta^2 - 2 (1-2\kappa) \lambda_{-} r_\Delta^4\big]
        }{
       4 v_{\rm ew} c_W^4  (1 + \lambda_{-} r_\Delta^2 )^2}  $ 
& 0 \\
$h W^-_\mu W^{+\mu} Z_\nu Z^\nu$ 
& $\displaystyle 
    \frac{
        \epsilon^2 g^4
        \big[(5-4\kappa)\kappa\, r_\Delta^2   - ( 4  - 11\kappa + 6\kappa^2) \lambda_{-} r_\Delta^4  \big]
        }{
        2 v_{\rm ew} c_W^2 (1 + \lambda_{-} r_\Delta^2 )^2 }$ 
& 0 \\
$h W^-_\mu W^+_\nu Z^\mu Z^\nu$
& $\displaystyle 
   -\frac{
        2\epsilon^2 g^4 
        \big[ 2 \big(  1 -c_W^2 - 4\kappa + 2 \kappa c_W^2  \big)  r_\Delta^2  +  (2-c_W^2)(1-2\kappa)\lambda_4 r_\Delta^4\big]
        }{
        v_{\rm ew}c_W^2 \big( 2 + (2\lambda_{-}+\lambda_5)  r_\Delta^2\big)^2} $ 
& 0 \\
$ihW^-_\mu W^+_\nu Z^{\mu\nu}$ 
& $\displaystyle 
    - \frac{\epsilon^2 g^3 (1-2\kappa)r_\Delta^2 }{v_{\rm ew} c_W \big( 2 +  (2\lambda_{-}+\lambda_5) r_\Delta^2 \big)}$ 
& 0 \\
$ (h i \overset{\leftrightarrow}{D}_\mu W^-_\nu) W^{+\mu} Z^\nu $  
& $\displaystyle 
    \frac{\epsilon^2 g^3 (1-2\kappa) r_\Delta^2 }{v_{\rm ew} c_W \big( 2 + (2\lambda_{-}+\lambda_5)r_\Delta^2  \big)}$ 
& 0 \\
$h^2\bar{\psi}_{L,p}\psi_{R,r}$ 
& $\displaystyle 
    \frac{
        \sqrt{2}\epsilon^2 \kappa
        (2-3\kappa +  \kappa\, \lambda_{-} r_\Delta^2 )
        }{
       v_{\rm ew} (1 + \lambda_{-} r_\Delta^2 ) }Y_\psi^{pr}$  
& $\displaystyle 
    \frac{3 \tilde{\epsilon}^2
    \big( 1 +2\lambda_{-}  \tilde{r}_\Delta^2 \big)}{\sqrt{2}\tilde{v}_{\rm ew} }  Y_\psi^{pr}$ \\
$W^-_\mu W^{+\mu} \bar{\psi}_{L,p}\psi_{R,r}$ 
& $\displaystyle 
    - \frac{
        \sqrt{2} \epsilon^2 g^2 r_\Delta^2\kappa(1-\kappa)
        }{
     v_{\rm ew} (1 + \lambda_{-} r_\Delta^2 ) } Y_\psi^{pr} $  
& $\displaystyle 0$ \\
$Z_\mu Z^\mu \bar{\psi}_{L,p}\psi_{R,r}$
& $\displaystyle 
    - \frac{
        \sqrt{2} \epsilon^2 g^2 (2-\kappa)\kappa r_\Delta^2
        }{
         2 v_{\rm ew}^{} c_W^2 (1 + \lambda_{-} r_\Delta^2 ) }  Y_\psi^{pr}  $  
& $\displaystyle 0$ \\
\hline
\multicolumn{3}{|c|}{$\Delta L =2$} \\
\hline
$h^2\l( \overline{\nu_{L,p}^{\tt C}} 
    \nu_{L,r} \r)$
& $\displaystyle 
    \frac{\epsilon ( 2-3\kappa + \kappa\, \lambda_{-} r_\Delta^2 )
    }{2 v_{\rm ew}  (1 + \lambda_{-} r_\Delta^2 )} Y_\Delta^{pr}  $ 
& $\displaystyle 
    -\frac{\tilde{\epsilon} }{2\tilde{v}_{\rm ew}^{}}
        (1 + \lambda_{-} \tilde{r}_\Delta^2) Y_\Delta^{pr}$ \\    
$W^-_\mu W^{+\mu} 
 \l( \overline{\nu_{L,p}^{\tt C}} 
    \nu_{L,r} \r)$ 
& $\displaystyle
    - \frac{\epsilon g^2  r_\Delta^2 (1-\kappa)
    }{ 2 v_{\rm ew}  (1 + 2 \lambda_{-} r_\Delta^2 ) } Y_\Delta^{pr}$ 
& 0 \\
$W^+_\mu W^{+\mu} \l( \overline{e_{L,p}^{\tt C}} e_{L,r} \r)$ 
& $\displaystyle 
    -\frac{\epsilon g^2  r_\Delta^2
    }{  2 v_{\rm ew} (1 + (\lambda_{-}+\lambda_5)r_\Delta^2) }  Y_\Delta^{pr}$ 
& 0 \\
$W^+_\mu Z^\mu \l( \overline{\nu_{L,p}^{\tt C}} e_{L,r} \r)$ 
& $\displaystyle 
    -\frac{\sqrt{2}\epsilon g^2 r_\Delta^2
    }{v_{\rm ew} c_W\big(2 + (2\lambda_{-}+\lambda_5) r_\Delta^2\big)} Y_\Delta^{pr} $ 
& 0 \\
$Z_\mu Z^\mu \l(  \overline{\nu_{L,p}^{\tt C}} \nu_{L,r} \r)$ 
& $\displaystyle 
    - \frac{\epsilon g^2  r_\Delta^2 (2-\kappa)
    }{2v_{\rm ew} c_W^2 (1 + 2 \lambda_{-} r_\Delta^2 )}  Y_\Delta^{pr}$ 
& 0 \\
\hline
\end{tabular}
}
\caption{ WCs for the dim-5 operators in the bEFT and bSMEFT. 
}
\label{table:heft-vs-smeft_5d}
\end{center}
\end{table}

\begin{table}[t]
\begin{center}
\renewcommand\arraystretch{1.6}
\begin{tabular}{|c|c||c|c|}
\hline
\rowcolor{lightgray} \multicolumn{4}{|c|}{\bf dim-6 with $\Delta L=0$} \\
\hline
Notation
& Operator
& WC in bEFT 
& WC in bSMEFT 
\\
\hline
${\cal O}_{(\partial h)^2 W^2}$
& $(\partial_\mu h)(\partial_\nu h)W^{-\mu}W^{+\nu}$ 
& $\displaystyle
    \frac{4\epsilon^2g^2 r_\Delta^2 (1-2\kappa)^2}{v_{\rm ew}^2 \big( 2 + (2\lambda_{-}+\lambda_5) r_\Delta^2 \big)}$ 
& 0 \\
${\cal O}_{h^2 (\partial W)^2}$
& $hh(D_\mu W^{-\mu}) (D_\nu W^{+\nu})$
& $\displaystyle
    -\frac{\epsilon^2 g^2 r_\Delta^2 (1-2\kappa)^2}{v_{\rm ew}^2 \big( 2 + (2\lambda_{-}+\lambda_5) r_\Delta^2 \big)}$ 
& 0 \\
${\cal O}_{h^2 \partial^2 WW}$
& $hh(D_\mu D_\nu W^{-\mu})W^{+\nu}$
& $\displaystyle
    -\frac{\epsilon^2 g^2 r_\Delta^2 (1-2\kappa)^2}{v_{\rm ew}^2 \big(2 + (2\lambda_{-}+\lambda_5) r_\Delta^2\big) }$ 
& 0 \\
${\cal O}_{(\partial h)^2 Z^2}$
& $(\partial_\mu h)(\partial_\nu h)Z^{\mu}Z^{\nu}$
& $\displaystyle
    \frac{2 \epsilon^2 g^2 r_\Delta^2 (1-2\kappa)^2}{v_{\rm ew}^2 c_W^2  (1 + \lambda_- r_\Delta^2 ) }$ 
& 0 \\
${\cal O}_{h^2 (\partial Z)^2}$
& $hh(\partial_\mu Z^\mu) (\partial_\nu Z^\nu)$
& $\displaystyle
    -\frac{\epsilon^2 g^2 r_\Delta^2 (1-2\kappa)^2}{2  v_{\rm ew}^2 c_W^2  (1 + \lambda_- r_\Delta^2 ) }$ 
& 0 \\
${\cal O}_{h^2 \partial^2 ZZ}$
& $hh Z^{\mu}(\partial_\nu \partial_\mu Z^{\nu})$ 
& $\displaystyle
    -\frac{\epsilon^2 g^2 r_\Delta^2 (1-2\kappa)^2}{v_{\rm ew}^2 c_W^2 (1 + \lambda_- r_\Delta^2 ) }$ 
& 0 \\
${\cal O}_{(\partial h)Zee}$
& $i(\partial_\mu h)Z^{\mu}(\bar{e}_R^p e_L^r)$ 
& $\displaystyle
    -\frac{2\sqrt{2}\epsilon^2 g r_\Delta^2 (2\kappa-1)}{v_{\rm ew}^2 c_W (1 + \lambda_- r_\Delta^2 ) }$ 
& 0 \\
${\cal O}_{h(\partial Z)ee}$
& $ih(\partial_\mu Z^{\mu})(\bar{e}_R^p e_L^r)$ 
& $\displaystyle
    -\frac{\sqrt{2}\epsilon^2 g r_\Delta^2 (2\kappa-1)}{v_{\rm ew}^2 c_W (1 +\lambda_- r_\Delta^2 ) }$ 
& 0 \\
\hline
\end{tabular}
\caption{
WCs for the dim-6 operators in the bEFT and bSMEFT.}
\label{table:heft-vs-smeft_6d}
\end{center}
\end{table}

\section{Numerical evaluation\label{sec:num}}
To assess how well the bEFT and bSMEFT frameworks reproduce the results of the UV theory, we calculate the cross sections for two typical processes involving the Higgs boson: 
one is the subprocess of Higgs pair production via vector boson fusion at the LHC, and the other is the Higgsstrahlung process at a future electron-positron collider. 
To calculate the scattering amplitude squared, we employ the Mathematica package {\tt FeynCalc} \cite{Shtabovenko:2016sxi,Mertig:1990an}. 
The original free parameters in the scalar potential are $m_H$, $\lambda$, $M_\Delta$, $\Lambda_6$, and $\lambda_i~(i=1,2,4,5)$.
After imposing minimization conditions, our input parameters are chosen as $\lambda$, $r_\Delta$, $\epsilon$, $v_{\rm ew}$, $\lambda_{-}$, and $\lambda_i~(i=1,2,5)$.
We set the scalar couplings $\lambda_1=\lambda_2=\lambda_5=1$ while varying the value of $\lambda_{-}$.
We checked that within the parameter region we focus on, the Higgs cubic, $hWW$, and $hZZ$ couplings satisfy the current LHC constraints, $-0.4<\kappa_\lambda<6.3$ \cite{ATLAS:2022jtk}, $-0.06 \leq \Delta\kappa_W \leq 0.10$, and $-0.03\leq \Delta\kappa_Z^{} \leq 0.11$ \cite{CMS:2022dwd}, respectively.
We assume the maximal value of $v_\Delta^{}\simeq 1\,{\rm GeV}$ that is allowed by the $\rho$ parameter, which implies $\epsilon \simeq 10^{-2}$. 
In this case, the doubly charged Higgs bosons $H^{\pm\pm}$ decay mainly into a pair of like-charge $W^\pm$ bosons, and the lower bound on their mass is relatively relaxed, $m_{H^{\pm\pm}}\gtrsim 400\,{\rm GeV}$ \cite{Chiang:2012dk,Ashanujjaman:2021txz}.

\begin{figure}
\centering
\includegraphics[width=0.45\textwidth,clip]{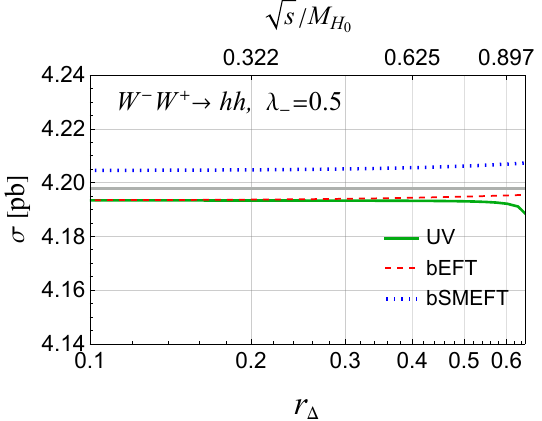} 
\hspace{5pt}
\includegraphics[width=0.45\textwidth,clip]{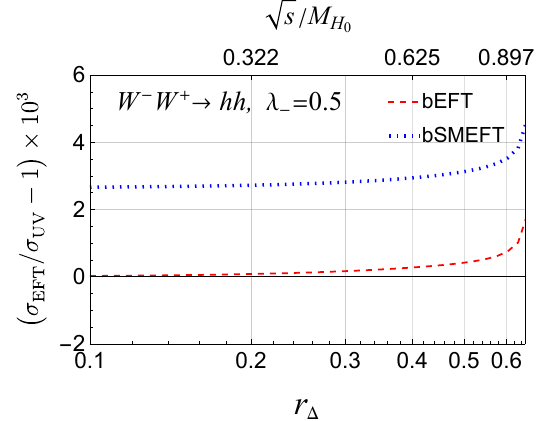} 
\vspace{10pt}
\\
\includegraphics[width=0.45\textwidth,clip]{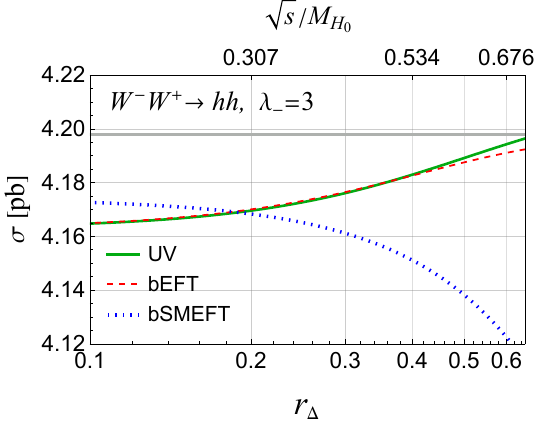} 
\hspace{5pt}
\includegraphics[width=0.45\textwidth,clip]{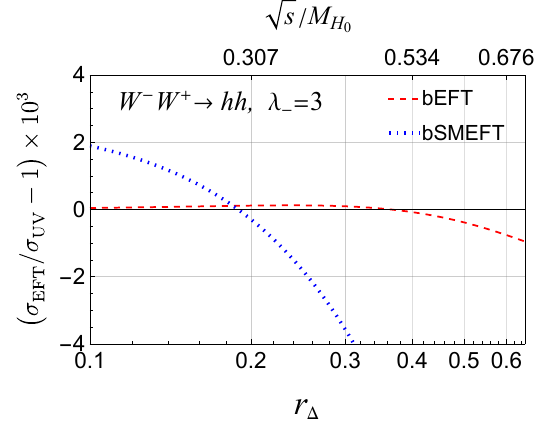} 
\vspace{10pt}
\\
\includegraphics[width=0.45\textwidth,clip]{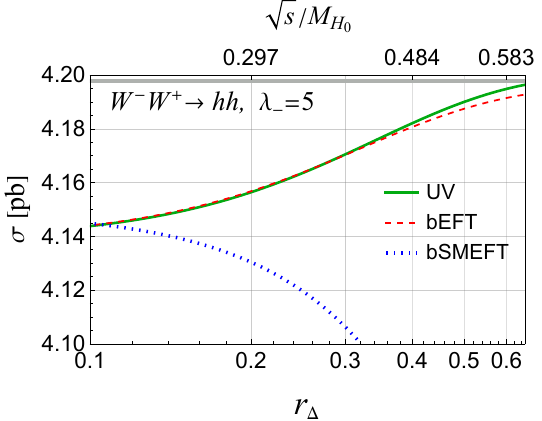} 
\hspace{5pt}
\includegraphics[width=0.45\textwidth,clip]{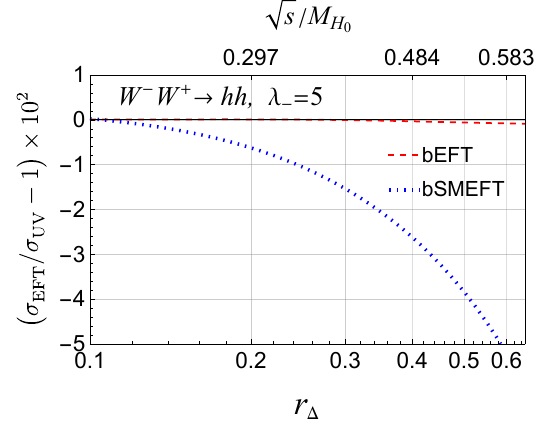} 
\caption{
Comparison of cross sections for $W^+W^-\to hh$ as a function of $r_\Delta$ or $\sqrt{s}/M_{H_0}$ calculated in the three theories at the fixed $\epsilon=10^{-2}$. 
The gray solid line on the left panels denotes the SM value.
}
\label{fig:WWhhrDelta}
\end{figure}

In \cref{fig:WWhhrDelta}, we show the cross section of $W^-W^+\to hh$ in each theory at a center-of-mass energy $\sqrt{s}=400~\rm GeV$ with a fixed value of $\epsilon=10^{-2}$.
We assume $\lambda_-=0.5$, $3$, and $5$ in the top, middle, and bottom panels, respectively.
In the left panels, the green solid, red dashed, and blue dotted curves correspond to the results in the UV theory, bEFT, and bSMEFT, while the gray solid line denotes the SM value.
In the right panels, we plot the relative deviations of cross sections in the bEFT and the bSMEFT from that in the UV. 
The upper horizontal axis in each plot gives the ratio of the center of mass energy to the mass of the lightest new scalar.
As can be seen from the figure, the bEFT reproduces very well the result of the UV theory in the whole range of the parameters shown. In contrast, the bSMEFT result deviates much more significantly from the UV result, and the deviation flips sign with the increase of $\lambda_-$.
As can be seen from the figure, even when we take the limit of $r_\Delta\to 0$, or equivalently, $M_\Delta\to \infty$, the UV and EFT cross sections still deviate from the SM value, indicating that the new physics effect does not decouple. This can be understood as follows: since we fix the $\epsilon$ to a finite value, the triplet VEV remains nonzero and it induces the new physics effect through $\lambda_{4,5}$ and $\Lambda_6$ interactions in \cref{eq:model_potential}.

\begin{figure}[htbp]
\centering
\includegraphics[width=0.49\textwidth,clip]{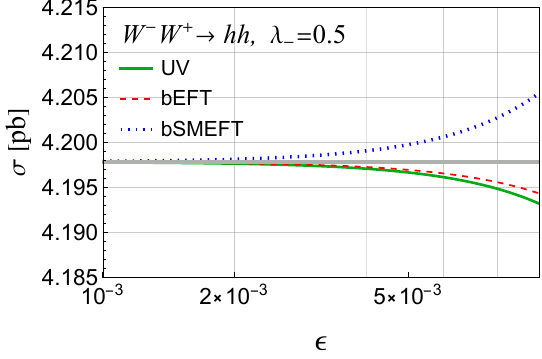} 
\hspace{5pt} 
\includegraphics[width=0.45\textwidth,clip]{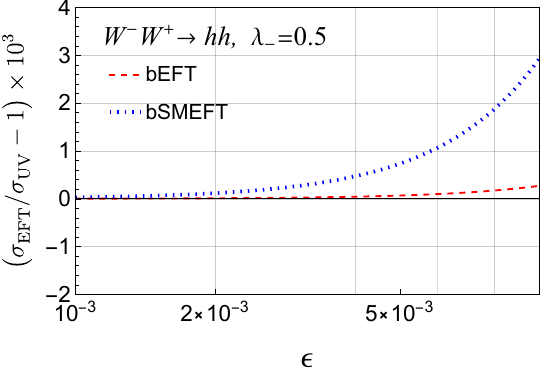} 
\vspace{5pt}
\\
\includegraphics[width=0.49\textwidth,clip]{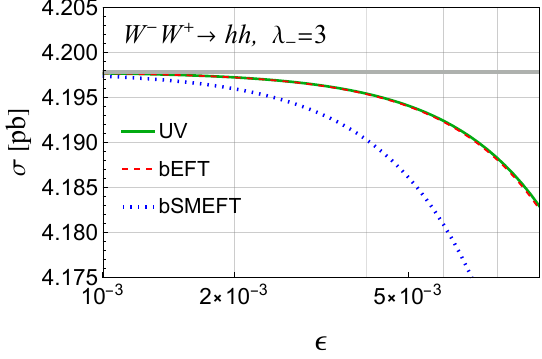} 
\hspace{5pt} 
\includegraphics[width=0.45\textwidth,clip]{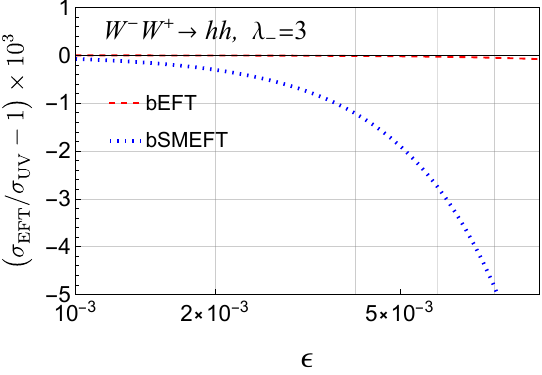} 
\vspace{5pt}
\\
\includegraphics[width=0.48\textwidth,clip]{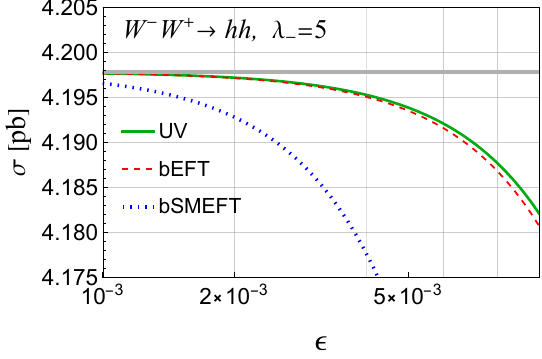} 
\hspace{5pt} 
\includegraphics[width=0.46\textwidth,clip]{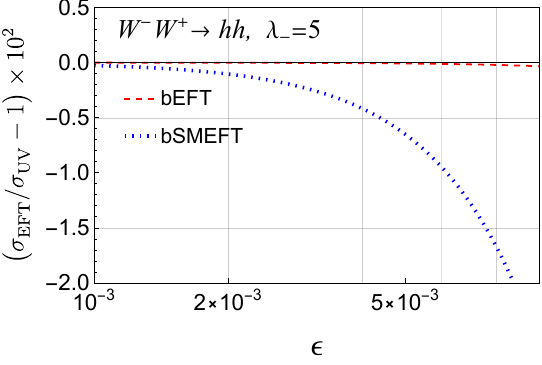} 
\caption{Comparison of cross sections for $W^+W^-\to hh$ as a function of $\epsilon$ calculated in the three theories at the fixed $r_\Delta=0.4$.
The gray solid line on the left panels denotes the SM value.}
\label{fig:WWhhepsilon}
\end{figure}

\begin{figure}[t]
\centering
\includegraphics[width=0.5\textwidth,clip]{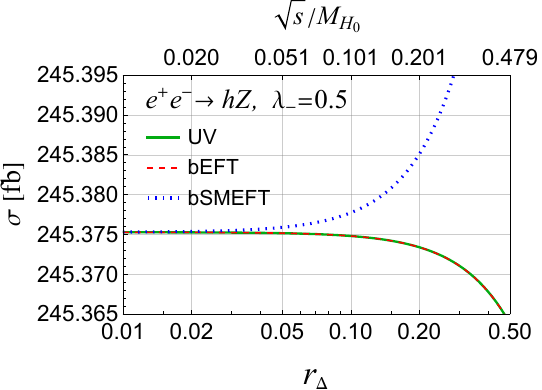} 
\hspace{5pt}
\includegraphics[width=0.45\textwidth,clip]{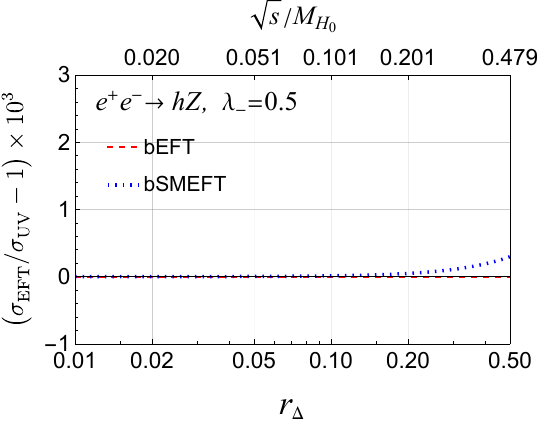} 
\vspace{10pt}
\\
\includegraphics[width=0.5\textwidth,clip]{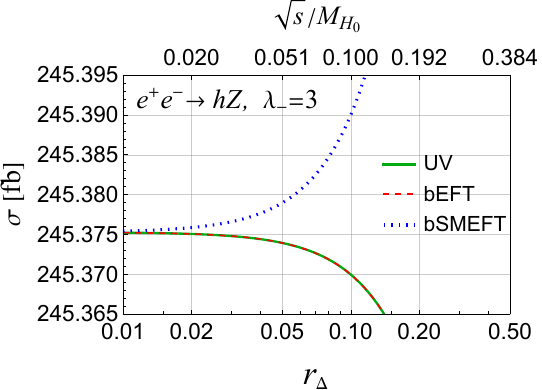} 
\hspace{5pt}
\includegraphics[width=0.45\textwidth,clip]{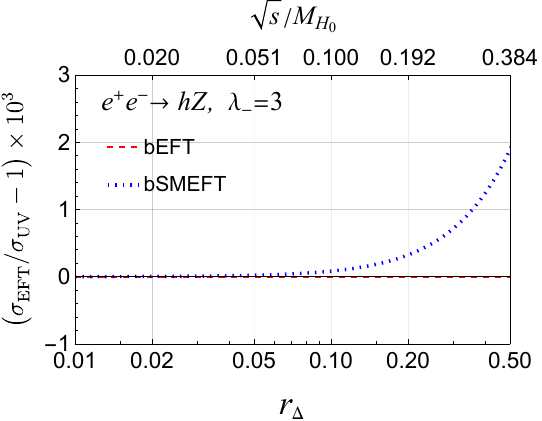} 
\vspace{10pt}
\\
\includegraphics[width=0.5\textwidth,clip]{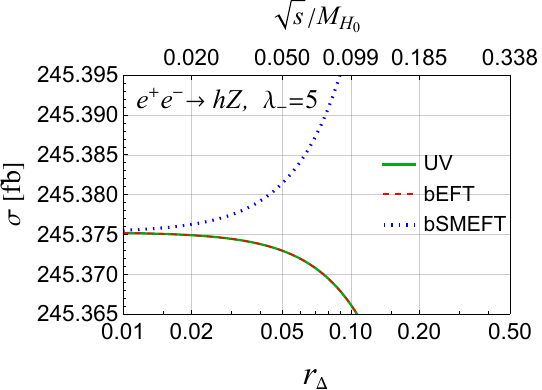} 
\hspace{5pt}
\includegraphics[width=0.45\textwidth,clip]{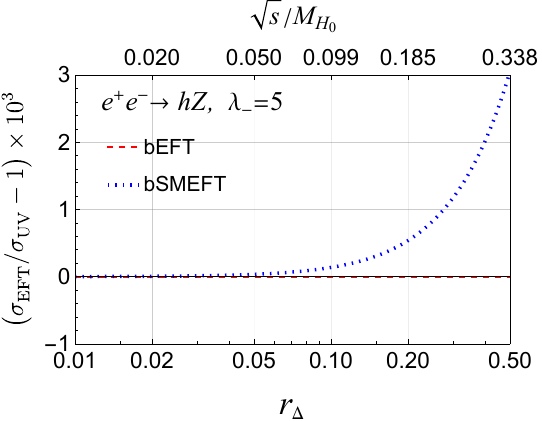} 
\caption{
Same as \cref{fig:WWhhrDelta} but for the process $e^+e^-\to hZ$.}
\label{fig:eehZrDelta}
\end{figure}

\begin{figure}[t]
\centering
\includegraphics[width=0.49\textwidth,clip]{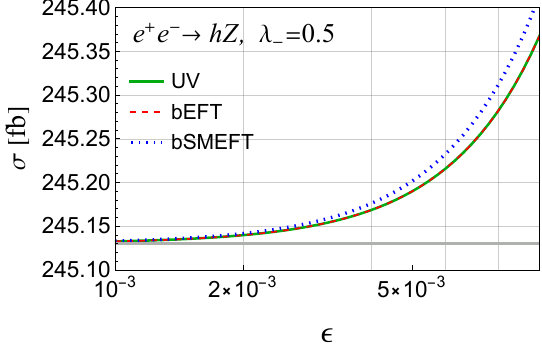} 
\hspace{5pt} 
\includegraphics[width=0.45\textwidth,clip]{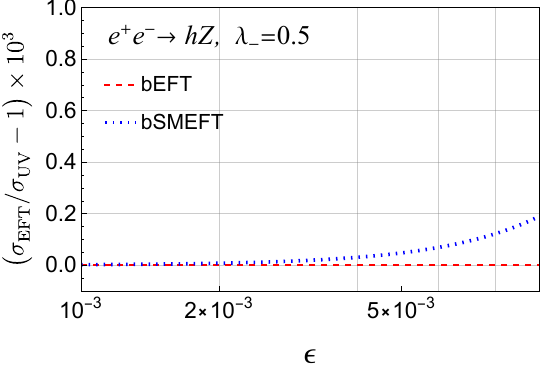} 
\vspace{5pt}
\\
\includegraphics[width=0.49\textwidth,clip]{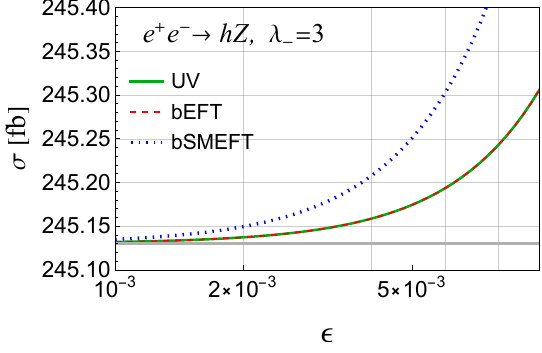} 
\hspace{5pt} 
\includegraphics[width=0.45\textwidth,clip]{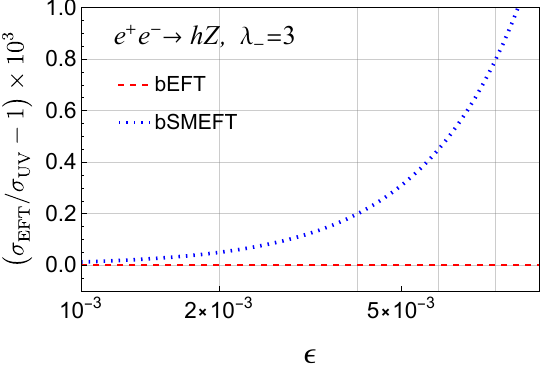} 
\vspace{5pt}
\\
\includegraphics[width=0.49\textwidth,clip]{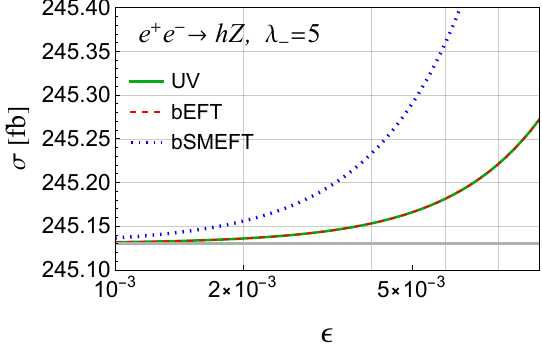} 
\hspace{5pt} 
\includegraphics[width=0.45\textwidth,clip]{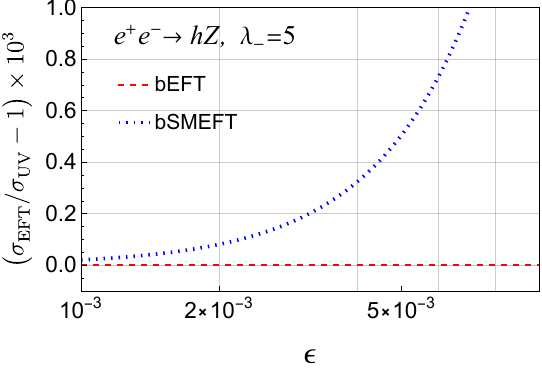} 
\caption{
Same as \cref{fig:WWhhepsilon} but for the process $e^+e^-\to hZ$.}
\label{fig:eehZepsilon}
\end{figure}

In \cref{fig:WWhhepsilon} the same cross section is plotted as a function of $\epsilon$ at the fixed $r_\Delta = 0.4$, using the same legends as in \cref{fig:WWhhrDelta}. 
Note that the masses of the heavy scalars vary with our input parameters. For instance, if we take the value of $\epsilon$ as $\epsilon=10^{-2}$, the masses of the heavy scalars are respectively, 
$(M_{H_0},M_{A_0},M_{H^\pm},M_{H^{\pm\pm}})/{\rm GeV}=(640,640,663,685)$ at $\lambda_{-}=0.5$, $(749,749,769,788)$ at $\lambda_{-}=3$, and $(826,826,844,862)$ at $\lambda_{-}=5$.
Again the bEFT result is precise at the per mille level or better while the bSMEFT deviates significantly from the UV theory for a slightly larger value of $\epsilon$.
As can be seen from the figure, when we take the limit of $\epsilon\to 0$ while fixing $r_\Delta=0.4$, the UV and EFT cross sections converge to the SM value. This reflects the following fact: the limit $\epsilon\to 0$ corresponds to the vanishing VEV of the scalar triplet, so it is equivalent to the limit of $\Lambda_6\to 0$. In this limit, the scalar triplet interacts with the Higgs doublet only in pairs in \cref{eq:model_potential}, so it does not affect $W^-W^+\to hh$ at tree level.
\cref{fig:eehZrDelta} and \cref{fig:eehZepsilon} show the corresponding results for the process
$e^+e^-\to Zh$ at $\sqrt{s}=250~\rm GeV$. 
The conclusion is the same as for the $W^-W^+\to hh$ process, with the main difference being that the bSMEFT result is constantly larger than the UV result.

The reason why the bEFT reproduces the results of the UV theory better than the bSMEFT is due to the difference in power counting between the two EFTs. 
We demonstrate this using the process $W^-(p_1)W^+(p_2)\to h(p_3)h(p_4)$ as an example.
As can be seen in \cref{table:heft-vs-smeft_6d}, the bEFT has dim-6 operators that contribute to $W^-W^+\to hh$, while the bSMEFT does not. 
This is because in the bSMEFT, effective operators with derivatives that give corrections to $W^-W^+\to hh$, such as $((D_\mu H)^\dag D_\mu H))^2$, only appear at the dim-8 level or higher.
This makes the bSMEFT less accurate in reproducing the UV results, as we show below.

In the UV theory, the contribution to $W^-W^+\to hh$ is given by
\begin{align}
\label{eq:vs_Muv}
\mathcal{M}^{\tt UV}
=
\mathcal{M}^{\tt UV}_h
+\mathcal{M}^{\tt UV}_{W^\pm}
+\mathcal{M}^{\tt UV}_{H_0}
+\mathcal{M}^{\tt UV}_{H^\pm}
+\mathcal{M}^{\tt UV}_{\rm contact} \, .
\end{align}
The terms $\mathcal{M}^{\tt UV}_{h,H_0}$ are the $s$-channel amplitudes mediated by $h,~H_0$, and the terms $\mathcal{M}^{\tt UV}_{W^\pm,H^\pm}$ are the $t,u$-channel amplitudes mediated by $W^\pm,~H^\pm$, respectively.
The term $\mathcal{M}^{\tt UV}_{\rm contact}$ is from the contact interaction.
In both EFTs, the contributions mediated by $H_0,H^\pm$ are incorporated into the contact part.
Note that $\mathcal{M}_{h,W^\pm}^{\tt UV}$ and $\mathcal{M}_{\rm contact}^{\tt UV}$ are different from their counterparts in the SM since the $hW^2$ and $h^2W^2$ couplings deviate from the SM through the scalar mixing.
Below we will show that the UV amplitude \eqref{eq:vs_Muv} coincides with the amplitude in the bEFT up to order $\epsilon^2$ if we take the limit of $s,~|t|,~|u|\ll M_\Delta^2$, where $s=(p_1+p_2)^2$, $t=(p_1-p_3)^2$, and $u=(p_1-p_4)^2$.

First of all, since the $h^3$ and $hW^2$ couplings in the UV theory coincide with those in the bEFT as we have emphasized in \cref{sec:vs}, the bEFT exactly reproduces the contributions $\mathcal{M}_{h,W^\pm}^{\tt UV}$. Consider next the contribution associated with $\mathcal{M}_{H^\pm}^{\tt UV}$. In the UV theory, it is  
\begin{align}
\label{eq:vs_uv-Hpm-mediated}
\mathcal{M}^{\tt UV}_{H^\pm}
&=
-(C^{\tt UV}_{hHW})^2
\l( \frac{(2p_3-p_1)_\mu(2p_4-p_2)_\nu}{t-m_{H^\pm}^2} +
    \frac{(2p_4-p_1)_\mu(2p_3-p_2)_\nu}{u-m_{H^\pm}^2} 
\r) 
\epsilon^\mu(p_1)
\epsilon^\nu(p_2) \, ,
\end{align}
where $\epsilon^\mu(p_i)$ is the polarization vector for the $W$ boson with momentum $p_i$. 
In the bEFT, on the other hand, it is replaced by the contact interactions $\mathcal{O}_{(\partial h)^2W^2}$, 
$\mathcal{O}_{h^2(\partial W)^2}$, and 
$\mathcal{O}_{h^2\partial^2 WW}$, resulting in 
\begin{align}
\label{eq:vs_beft-dim6}
\mathcal{M}^{\tt bEFT}_{{\rm dim}\mathchar`-6} 
&=
\frac{\epsilon^2 g^2r_\Delta^2 (1-2\kappa)^2}{v_{\rm ew}^2(2+r_\Delta^2\lambda_4)}
\Big[(2p_3-p_1)_\mu(2p_4-p_2)_\nu +(2p_4-p_1)_\mu(2p_3-p_2)_\nu\Big]
\epsilon^\mu(p_1)\epsilon^\nu(p_2) \, .
\end{align}
\cref{eq:vs_uv-Hpm-mediated} coincides with \cref{eq:vs_beft-dim6} up to order $\epsilon^2$ in the limit of $|t|,~|u|\ll m_{H^\pm}^2$ by the relation, 
\begin{align}
\frac{(C_{hHW}^{\tt UV})^2}{m_{H^\pm}^2}
=
\frac{\epsilon^2 g^2r_\Delta^2 (1-2\kappa)^2}{v_{\rm ew}^2(2+r_\Delta^2\lambda_4)}
+
\mathcal{O}(\epsilon^4) \, .
\end{align}

Lastly, we show how the bEFT reproduces $\mathcal{M}^{\tt UV}_{H_0}$ and $\mathcal{M}^{\tt UV}_{\rm contact}$. Their explicit forms are given by
\begin{subequations}
\begin{align}
&\mathcal{M}_{H_0}^{\tt UV}
=
-C_{h^2H_0}^{\tt UV}C_{H_0W^2}^{\tt UV}\frac{1}{s-M_{H_0}^2} \, ,
\\
&\mathcal{M}^{\tt UV}_{\rm contact} = C_{h^2W^2}^{\tt UV} \, .
\end{align}
\end{subequations}
In the bEFT, they are replaced by the contact interaction $h^2W_\mu^-W^{+\mu}$ given in \cref{table:heft-vs-smeft_4d}:
\begin{align}
\label{eq:vs_beft-dim4}
\mathcal{M}^{\tt bEFT}_{\rm dim\mathchar`-4}
=
\frac{g^2}{4}\l[ 1 - \epsilon^2\frac{4(1-2\kappa)(2-\kappa+\kappa\lambda_{-}r_\Delta^2)}{1+\lambda_{-}r_\Delta^2} \r] \, .
\end{align}
It is easy to see that the sum of $\mathcal{M}^{\tt UV}_{H_0}$ and $\mathcal{M}^{\tt UV}_{\rm contact}$ coincides with \cref{eq:vs_beft-dim4} up to order $\epsilon^2$  in the limit of $s\ll m_{H_0}^2$ by applying the relation, 
\begin{align}
\frac{C_{h^2H_0}^{\tt UV}C_{H_0W^2}^{\tt UV}}{M_{H_0}^2}
+C_{h^2W^2}^{\tt UV}
=
\mathcal{M}^{\tt bEFT}_{\rm dim\mathchar`-4} \, + \mathcal{O}(\epsilon^4) \, .
\end{align}
To summarize, the UV amplitude \eqref{eq:vs_Muv} coincides with the amplitude in the bEFT up to order $\epsilon^2$ in the low energy limit of $s,~|t|,~|u|\ll M_\Delta^2$.

In the bSMEFT, the heavy fields are integrated out before the electroweak symmetry breaking. 
The drawback with this is that the bSMEFT does not properly take into account the effects of the scalar mixing in the broken phase unless we work to a higher order in power counting.
For example, the coupling $C_{hHW}^{\tt UV}$ appearing in $\mathcal{M}_{H^\pm}^{\tt UV}$ is generated by the mixing between the charged components in the ${\rm SU(2)_L}$ doublet $H$ and triplet $\Delta$. 
In the bEFT, the scalar mixing is incorporated at the very start, so the contribution of \cref{eq:vs_beft-dim6} arises from the dim-6 operators and reproduces $\mathcal{M}_{H^\pm}^{\tt UV}$ in the low energy limit. 
In contrast, there is no contact interaction with derivatives in the bSMEFT up to the dim-6 level that contributes to $W^-W^+\to hh$, as can be seen in \cref{table:heft-vs-smeft_6d}. 
Although the bSMEFT has an $h^2W^2$ operator as shown in \cref{table:heft-vs-smeft_4d}, this operator does not involve derivatives and therefore is not enough to describe accurately the momentum dependence of processes mediated by $H^\pm$. As a matter of fact, 
the bSMEFT treats the $H_0$, $A_0$, $H^\pm$, and $H^{\pm\pm}$ together as a single triplet field. This means it makes not much sense to speak of their separate contributions at the leading low energy order.
Now the reason becomes clear why the bEFT reproduces the low energy result of the UV theory more accurately than the bSMEFT: it comes from the difference in power counting. The bEFT incorporates the effects of scalar mixing at low orders of the perturbative expansion, while the bSMEFT cannot unless it is expanded to higher orders. 
Therefore the bEFT can be considered as a better organization of low energy expansion of the UV theory.
On the other hand, although we have not performed the matching of the type-II seesaw onto the HEFT for comparison, this procedure is expected to be more complicated due to the matrix representation of the Goldstone bosons, as well as the intricate field redefinitions and power counting involved. Nevertheless, we expect both the bEFT and HEFT to exhibit similar convergence in reproducing the model's predictions.

Before closing this section, we briefly comment on applying the bEFT at the loop level. Although the bEFT is derived at the tree level in the unitary gauge, as exemplified here by the type-II seesaw, it can be consistently extended to the loop level and with other gauge choices. 
In all cases, gauge invariance in the UV theory is still maintained, albeit not in a manifest manner. In principle, loop-level matching calculations can be performed in the unitary gauge without introducing inconsistencies, but at the expense of poor UV behavior which has to be handled with care. 
On the other hand, to make gauge invariance more explicit and have better UV behavior in loop calculations, one may choose a different gauge that retains the would-be Goldstone bosons as relevant degrees of freedom. In this approach, the bEFT explicitly includes the Goldstone bosons associated with SM gauge bosons, and consequently, the number of relevant EFT operators tends to increase with the operator dimension.

\section{Conclusion\label{sec:concl}}

In this work we have introduced the broken phase EFT (bEFT) whose symmetry is $\rm SU(3)_c\times U(1)_{em}$ and degrees of freedom are the SM particles in the broken phase.
We considered the matching of the type-II seesaw model onto the bEFT at tree level up to dim-6 operators. 
To make comparisons with the SMEFT, we rewrote the results of matching the type-II seesaw model onto the SMEFT in a basis after the symmetry breaking, named as bSMEFT. 
To compare the two EFTs concretely, we evaluated the Higgs pair production through vector-boson fusion and the single Higgs production through the Higgsstrahlung process using the type-II seesaw, bEFT, and bSMEFT. 
We found that the bEFT generically reproduces the results of the type-II seesaw model more accurately than the bSMEFT. 
The difference arises from power counting, i.e., the organization of perturbative expansion.
In the bSMEFT, the heavy fields are integrated out before electroweak symmetry breaking, so that the effect of mixing between the scalar fields is not incorporated at the leading order.
In the bEFT, on the other hand, the heavy fields are integrated out after electroweak symmetry breaking whence their mixing with the light scalar has been taken into account. 
This is in accord with the physical intuition that the SMEFT may not work well when new heavy scalars participate in the electroweak symmetry breaking or when new particles are not heavy enough compared with the electroweak scale. 
Therefore, we expect the bEFT to provide a more accurate EFT description the SM is extended by singlet or triplet scalars, as well as in the two-Higgs-doublet model.
Furthermore, the bEFT can provide a suitable framework for UV models with extended gauge groups, such as the left-right symmetric model \cite{Mohapatra:1979ia}, the Pati-Salam model \cite{Pati:1974yy}, and grand unified theories. In these models, the SM fields intertwine with new heavy fields in specific representations. One must go into the broken phase, identify the SM particles and heavy new particles, and integrate out the latter to obtain the bEFT. We will apply the bEFT to these cases in our future work.

\section*{Acknowledgements}
YU would like to thank Masaharu Tanabashi 
and Koji Tsumura for the fruitful discussion.
This work was supported in part by the Grants 
No.\,NSFC-12305110, 
No.\,NSFC-12035008, 
and No.\,NSFC-12347112, 
and by the Guangdong Major Project of Basic
and Applied Basic Research No.\,2020B0301030008.


\appendix
\setcounter{section}{0}
\section{Minimization of the scalar potential}
\label{app:minimization}

We derive the minimization condition of the scalar potential in \cref{eq:model_potential} and the mass eigenstates of the scalar fields after symmetry breaking.
Substituting the VEVs of the ${\rm SU(2)_L}$ doublet and
triplet scalar fields denoted by 
$\langle H\rangle$ and $\langle \Delta\rangle$
into the scalar potential, 
we get
\begin{align}
    \label{eq:model_potential-vev}
    &\mathcal{V}(\langle H\rangle, \langle\Delta\rangle)
    =
    \frac{1}{2}M_{\Delta}^2v_\Delta^2
    +\frac{1}{8}\lambda v_H^4
    +\frac{1}{8}\lambda_1 v_\Delta^4
    +\frac{1}{2}\lambda_{-} v_H^2v_\Delta^2 
    -\frac{1}{2}\Lambda_6v_H^2 v_\Delta
    -\frac{1}{2}m_H^2v_H^2 \, .
\end{align}
Note that $v_H^{}$ and $v_\Delta^{}$
satisfy the following relation,
\begin{align}
\label{eq:model_vT}
v_{\rm ew}^{} = \sqrt{v_H^2 + 2v_\Delta^2} \, ,
\end{align}
with $v_{\rm ew} = 246$ GeV.
The conditions for the scalar potential to take an extreme value 
at $v_H$ and $v_\Delta$ are 
\begin{subequations}
\begin{align}
    \label{eq:model_minimal}
    &\frac{\partial \mathcal{V}(\langle H\rangle, \langle\Delta\rangle)}{\partial v_H }
    =
    v_H^{}
    \left( \frac{1}{2}\lambda v_H^2 
           -\Lambda_6 v_\Delta
           +\lambda_{-}v_\Delta^2
           -m_H^2 
    \right) 
    =
    0 \, ,
    \\
    &\frac{\partial \mathcal{V}(\langle H\rangle, \langle\Delta\rangle)}{\partial v_\Delta}
    =
    v_\Delta
    \left( -\frac{1}{2}\frac{ \Lambda_6v_H^2}{v_\Delta}
           +\lambda_{-}v_H^2
           +\frac{1}{2}\lambda_1v_\Delta^2
           +M_\Delta^2 
    \right) 
    =
    0 \, .       
\end{align}
\end{subequations}
Solving the above, we get
\begin{subequations}
\begin{align}
    \label{eq:model_minimization-1}
    &m_H^2
    =
    \frac{1}{2}\lambda v_H^2 
    -\Lambda_6 v_\Delta
    +\lambda_{-}v_\Delta^2 \, ,
    \\
    \label{eq:model_minimization-2}
    &M_\Delta^2
    =
    \frac{1}{2}\frac{\Lambda_6v_H^2}{v_\Delta}
    -\lambda_{-}v_H^2
    -\frac{1}{2}\lambda_1v_\Delta^2 \, .
\end{align}
\end{subequations}
We replace the original parameters $m_H^2$ and $\Lambda_6$ 
with the two VEVs by using the above two relations. 
The mass terms for the scalar fields are given by
\begin{align}
 -   \mathcal{L}_{\rm mass}
    =&\,
    \frac{1}{2}(\,h, ~\delta^0\,)
    ~\mathcal{M}_{\rm CPeven}^2
    \left( 
    \begin{array}{c}
        h
        \\
        \delta^0 
    \end{array}
    \right)
    +
    \frac{1}{2}(\,\chi, ~\eta\,)
    ~\mathcal{M}_{\rm CPodd}^2
    \left( 
    \begin{array}{c}
        \chi
        \\
        \eta
    \end{array}
    \right)
    \nonumber\\
    &\,
    +(\,H^-, ~\delta^-\,)
    ~\mathcal{M}_{\rm Charged}^2
    \left( 
    \begin{array}{c}
        H^+
        \\
        \delta^+ 
    \end{array}
    \right)
    +
    m_{H^{\pm\pm}}^2H^{--}H^{++} \, ,
\end{align}
where the explicit forms of mass matrices are 
\begin{subequations}
\begin{align}
    \label{eq:model_MsqCPeven}
    &\mathcal{M}_{\rm CPeven}^2
    =
    \left( 
    \begin{array}{cc}
        \lambda v_H^2 & -\frac{v_\Delta}{v_H^{}} (2M_\Delta^2 + \lambda_1v_\Delta^2) 
        \\
        -\frac{v_\Delta}{v_H^{}} (2M_\Delta^2 + \lambda_1v_\Delta^2) & M_\Delta^2+\frac{3}{2}\lambda_1v_\Delta^2 + \lambda_{-}v_H^2 
    \end{array}
    \right) \, ,
    \\
    \label{eq:model_MsqCPodd}
    &\mathcal{M}_{\rm CPodd}^2
    =
    \frac{2M_\Delta^2 + \lambda_1v_\Delta^2 + 2\lambda_{-}v_H^2}{2v_H^2}
    \left( 
    \begin{array}{cc}
        4v_\Delta^2 & -2v_H^{}v_\Delta
        \\
        -2v_H^{}v_\Delta & v_H^2 
    \end{array}
    \right) \, ,
    \\
    \label{eq:model_MsqCharged}
    &\mathcal{M}_{\rm Charged}^2
    =
    \frac{2M_\Delta^2 + \lambda_1v_\Delta^2 + \lambda_4v_H^2}{2v_H^2}
    \left( 
    \begin{array}{cc}
        2v_\Delta^2 & -\sqrt{2}v_H^{}v_\Delta
        \\
        -\sqrt{2}v_H^{}v_\Delta & v_H^2 
    \end{array}
    \right) \, .
\end{align}
\end{subequations}
By diagonalizing the above mass matrices,
we obtain the masses for the scalar fields as
\begin{subequations}
\begin{align}
    &m_{H^{\pm\pm}}^2
    =
    M_\Delta^2
    + (\lambda_{-}+\lambda_5) v_H^2
    +\frac{1}{2}(\lambda_1+\lambda_2)v_\Delta^2 \, ,
    \\
    &m_{H^\pm}^2
    =
    \bigg( M_\Delta^2 
           +\frac{1}{2}\lambda_4 v_H^2
           +\frac{1}{2}\lambda_1 v_\Delta^2
    \bigg)
    \bigg(
    1
    +\frac{2v_\Delta^2}{v_H^2}
    \bigg) \, ,
        \\
    &m_{A_0}^2
    =
    \bigg( M_\Delta^2
           +\lambda_{-}v_H^2
           +\frac{1}{2}\lambda_1 v_\Delta^2
    \bigg)
    \bigg(
    1
    +\frac{4v_\Delta^2}{v_H^2}
    \bigg) \, ,
    \\
    \label{eq:model_mh}
    &m_{h}^2
    =
    \frac{1}{2}
    \Big( A+C-\sqrt{(A-C)^2+4B^2} \Big) \ ,
    \\
    \label{eq:model_mH0}
    &m_{H_0}^2
    =
    \frac{1}{2}
    \Big( A+C+\sqrt{(A-C)^2+4B^2} \Big) \ ,
\end{align}
\end{subequations}
where $A$, $B$, and $C$ are given by
\begin{subequations}
\begin{align}
    &A = \lambda v_H^2 \, ,
    \\
    &B = -\frac{2v_\Delta}{v_H^{}} 
          \bigg( M_\Delta^2 + \frac{1}{2}\lambda_1v_\Delta^2 \bigg) \, ,
    \\
    &C = M_\Delta^2 
         + \lambda_{-}v_H^2 
         + \frac{3}{2}\lambda_1 v_\Delta^2 \, .
\end{align}
\end{subequations}

\bibliography{references.bib}{}
\bibliographystyle{utphys28mod}

\end{document}